RESEARCH

# Improving radial velocity precision with CARMENES-PLUS

## An upgrade of the near-infrared spectrograph cooling system

R. Varas[1] · R. Calvo-Ortega[1] · P. J. Amado[1] · S. Becerril[2] · H. Ruh[3] · M. Azzaro[4] · L. Hernández[4] · H. Magán-Madinabeitia[1,4] · S. Reinhart[4] · D. Maroto-Fernández[4] · J. Helmling[4] · Á. L. Huelmo[1] · D. Benítez[4] · J. F. López[4] · M. Pineda[4] · J. A. García[4] · J. García de la Fuente[4] · J. Marín[4] · F. Hernández[4] · J. Aceituno[4] · J. A. Caballero[5] · A. Kaminski[6] · R. J. Mathar[7] · A. Quirrenbach[6] · A. Reiners[3] · I. Ribas[8,9] · W. Seifert[6] · M. Zechmeister[3]

**Abstract**

**Context:** CARMENES is a dual-channel high-resolution spectrograph at the 3.5 m Calar Alto telescope designed to detect low-mass planets around late-type dwarfs by measuring their radial velocities (RVs). High thermal stability in both the visible (VIS) and near-infrared (NIR) channels is essential to achieve the precision required for these measurements. In particular, stabilising the NIR channel to the millikelvin level, which operates at cryogenic temperatures ( $\sim$ 140 K), poses significant engineering challenges. **Purpose:** The CARMENES-PLUS project was initiated to improve the instrument's intrinsic RV precision. In this article, we focus on the thermal stability improvements made to the NIR channel's cooling system. **Methods:** The NIR cooling system was originally conceived to operate with a discontinuous flow of cryogenic nitrogen gas. As part of CARMENES-PLUS, this was upgraded to a continuous flow configuration. Additional changes included the installation of an automatic vacuum system, a proportional control valve, and a pressure regulation system. These upgrades were designed to reduce thermal fluctuations and enhance long-term stability. **Results:** The implemented upgrades significantly improved the intrinsic RV precision of the NIR channel. We quantified this improvement using Fabry-Pérot calibration spectra, obtaining an intrinsic RV precision of 0.67 m $s^{-1}$ after the interventions, an improvement of nearly 2 m $s^{-1}$. We also assessed the stability of the nightly zero points, finding a reduced scatter of 3.9 m $s^{-1}$ post-upgrade, compared to 6.1 m $s^{-1}$ before. For a sample of slowly rotating stars ($v \sin i_\star \leq 2$ km $s^{-1}$), the median scatter decreased from 8.8 m $s^{-1}$ to 6.7 m $s^{-1}$ after the upgrades. **Conclusions**: These results demonstrate that the thermal control upgrades introduced in CARMENES-PLUS have enhanced the NIR channel's RV performance, bringing it closer to the VIS channel's stability and reinforcing CARMENES's capabilities for exoplanet detection around M dwarfs.

Extended author information available on the last page of the article

## 1 Introduction

Astronomical instruments operating in the near-infrared (NIR) or at longer wavelengths require cooling to reduce the environmental thermal background, which would otherwise dominate the detector signal and degrade performance. In recent years, a new generation of high-resolution spectrographs working at NIR wavelengths has been developed specifically to deliver precise radial velocity (RV) measurements of late-type dwarfs, key targets in the search for low-mass exoplanets.. Notable examples include GIANO-B [1], Infrared Doppler instrument [IRD, 2], Habitable zone Planet Finder [HPF, 3], SPectropolarimètre InfraROUge [4, SPIRou], Near Infra Red Planet Searcher [5, NIRPS], CRIRES+ [6], and CARMENES [7].

CARMENES[1] is a dual-channel high-resolution spectrograph installed at the 3.5 m telescope at the Calar Alto Observatory (CAHA, Almería, Spain). It was one of the first near-infrared spectrographs to operate routinely at a telescope and the first to include both optical and near-infrared arms (VIS: 520–960 nm; NIR: 960–1710 nm) for simultaneous observations.

Since the start of its operation in 2016, the instrument has been used to fulfil its main science objective, detecting low-mass planets in the habitable zones of a sample of 300 nearby bright M-dwarf stars [8]. The VIS channel has been used for characterising these planetary systems, including their host stars [9–12] whereas the NIR channel made a breakthrough in the study of planetary atmospheres with the detection of Helium escape from hot giant planets [13, 14]. Simultaneous data from both channels have been used together for planetary atmospheric analysis or stellar characterisation [15, 16], but only very few times for RV studies [17], mainly of massive planets.

GIANO-B and HARPS-N [1] and NIRPS and HARPS [18] are pairs of near-infrared and optical high-resolution spectrographs that can operate simultaneously at the Telescopio Nazionale Galileo and the ESO 3.6 m La Silla telescope, respectively. The two telescopes have aperture sizes similar to the 3.5 m Calar Alto telescope. However both pairs of spectrographs were not originally designed to operate together, which affects their operational integration, and have a large wavelength gap between 690 nm and 950 nm, here most of the RV information content is located in M dwarfs [19]. CARMENES, having been designed specifically for planet searches around M dwarfs, does not have any major wavelength gap between 520 nm and 1710 nm, and their two channels were designed for simultaneous and integrated operation.

To extract the full potential of both channels, achieving the best possible intrinsic stability is essential. For the NIR channel, this is in part achieved by a high thermal stability.

[1] Calar Alto high-Resolution search for M dwarfs with Exoearths with Near-infrared and optical Echelle Spectrographs, https://carmenes.caha.es

There are several solutions for providing the stable cooling needed by these NIR instruments, from the standard solution used for detectors, which implements the liquid nitrogen continuous flow cryostat as those developed by ESO ([20]), to the off-the-shelf solution provided by cryocoolers [21]. Most of these instruments operate under full cryogenic conditions (i.e., at the temperature of liquid nitrogen -LN2- boiling point, 77 K), which poses technical challenges when there are movable parts (e.g. EMIR; [22]). A less common solution is evaporating the LN2 to generate a flow of nitrogen gas at intermediate cryogenic temperatures above the LN2 boiling point, and feed it through the instrument to cool it down. This is the solution chosen for cooling the CARMENES NIR channel.

Bauer et al. [23] carried out the first thorough analysis of the RV performance of the two CARMENES channels during the first years of operation. They confirmed the higher intrinsic RV scatter of the the NIR channel, 3.7 m $s^{-1}$, compared to 1.2 m $s^{-1}$ in the VIS, due to limitations in calibration precision and thermal stability. Thus, there was room for improving the intrinsic RV precision in the NIR channel.

To improve the instrument's technical performance, the project CARMENES-PLUS (C-PLUS) was born, establishing a plan of upgrades for the different CARMENES subsystems. This work focuses on the improvements implemented in the NIR cooling system (CS).

The outline of this article is as follows: in Sect. 2 we provide an overview of the CS of the CARMENES NIR spectrograph as it was designed in first place. In Sect. 3 we give the list and the description of the upgrades carried out in the NIR spectrograph within the C-PLUS project along with the new thermal stability. This paper focuses on the evaluation of the thermal improvements introduced in the NIR channel of CARMENES through the C-PLUS upgrade. Sect. 4 describes the data reduction and calibration strategies, including Fabry-Pérot calibration spectra, RV-constant stars, and M-dwarf stellar observations. In Sect. 5, we analyze these datasets to quantify the improvement in instrumental stability, nightly zero point scatter, and overall RV precision for scientific targets. These results demonstrate the relevance of the thermal control upgrades for enhancing the scientific capabilities of the instrument. Finally, we summarise the main conclusions in Sect. 6.

## 2 CARMENES NIR spectrograph cooling system (2016-2020)

The CARMENES NIR spectrograph requires a CS (Fig. 1) in order to maintain all its optical elements at cryogenic temperature [24, 25]. Its opto-mechanics (all the optical elements and their mounts) are attached to a massive optical bench, which is surrounded by a radiation shield with a 20-layered multi-layer insulation blanket. Both the optical bench and the radiation shield are enclosed inside a vacuum tank, making convection heat transfer and the fluctuations of the refractive index of the residual air negligible. The nitrogen gas flows through pipes that reach heat exchangers on the radiation shield, dissipating the radiative load coming from the vacuum tank to keep the radiation shield at working temperature ($\sim$ 140 K). The vacuum tank is kept in an isolated and thermally stabilised room which in turn is located at the Coudé room of the telescope, both maintained at 12 ± 0.1 °C.

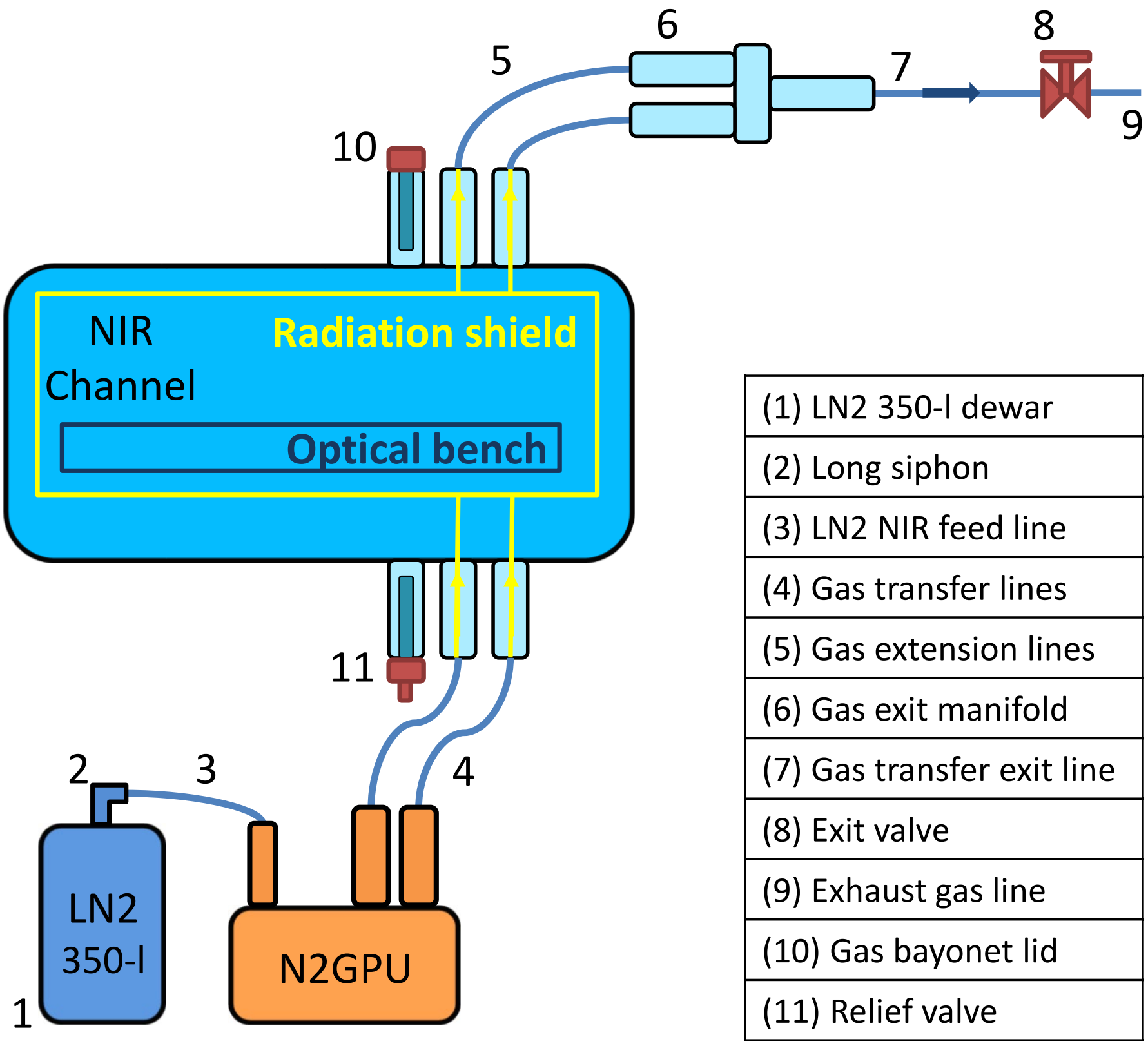

**Fig. 1** Simplified layout of the CARMENES cooling system during normal operation showing the main components, as described in [25]

The radiation shield passively thermalises the optical bench via the radiation heat transfer mechanism. The conduction heat transfer between the cold parts (the optical bench and the radiation shield) and the vacuum tank is minimised by using G-10 epoxy fibreglass and appropriate interface geometries.

A separate nitrogen gas preparation unit (N2GPU) supplies the nitrogen gas to the radiation shield. This N2GPU is a prototype of ESO's standard continuous flow cryostats, adapted to apply this concept to a spectrograph's CS. Although designed for high thermal stability, in practice it did not achieve the expected performance. A 350-l dewar feeds LN2 into this N2GPU. This dewar is replaced by a full and pressurised one every morning. The N2GPU is equipped with several stages that evaporate the LN2 and bring the gas to the working temperature using two proportional–integral–derivative control loops. It was designed to provide a highly thermally stable coolant, eliminating the need for active thermal control heaters inside the vacuum tank and, consequently, removing the system's dependence on their performance and risk of failure.

Just at the exit of the N2GPU, the coolant is split into two independent circuits, flowing through two vacuum insulated gas transfer lines. The coolant enters the vac-uum tank through a pair of flow splitters that divide the two circuits into ten cooling lines, feeding a total of 19 heat exchangers that cool the radiation shield. The warm gas is evacuated through a single return circuit.

Originally, the thermalisation was performed by a control loop with a cryogenic on/off valve installed in the exit circuit, similarly to ESO's standard continuous-flow cryostats. The loop controls the valve by comparing the temperature measured by the reference sensor on the radiation shield to a predefined setpoint. The open/close states of the on/off valve kept the temperature value within a dead-band below the setpoint: when the temperature sensor value reached the setpoint, the on/off valve opened to cool down the radiation shield, whereas the warming phase occurred when the valve was closed, allowing the coolant to remain in the circuit.

The objective of this design and thermal stabilisation strategy was to use the high thermal inertia of the metallic mass of the optical bench to dampen temperature variations from the radiation shield, ensuring that the opto-mechanics remained thermally stable.

Short-term variations in the shield introduced by the on/off valve (see Fig. 2) were anticipated and expected to be suppressed by the bench's damping effect. However, in practice, these fluctuations–together with additional variability caused by the underperforming N2GPU–were not fully absorbed and were transmitted to the opto-mechanics. As a result, the RVs measured with the NIR channel showed a direct correlation with the thermal variability of the radiation shield, through a mechanism that remains not fully understood.

This discovery indicated that any improvement in the thermal stability of the spectrograph, and hence the intrinsic precision of its RVs, required a reduction of the thermal variability of the radiation shield and, consequently, an enhancement of the N2GPU's thermal stabilisation performance.

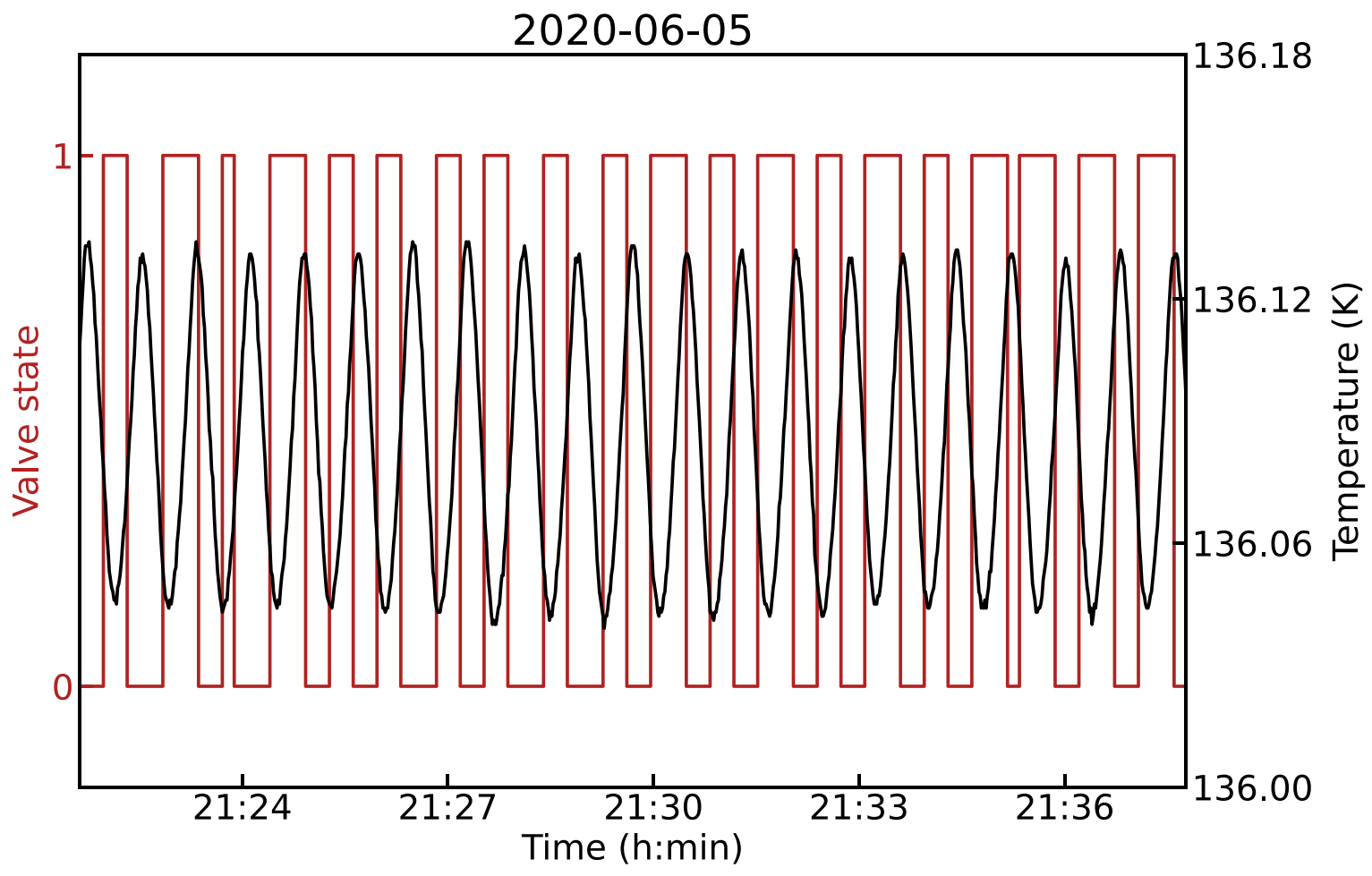

**Fig. 2** Temperature response of the radiation shield (black) to the binary states of the on/off valve (red), recorded on 2020-06-05. The valve opens and closes periodically, producing a sawtooth temperature pattern characteristic of the discontinuous flow mode

## 3 CARMENES-PLUS: A step further to enhance the NIR thermal stability

The CARMENES-PLUS (C-PLUS) project was proposed to maintain CARMENES at the forefront of technical performance by improving its intrinsic RV precision to 2 $m\ s^{-1}$ over a timescale longer than five years and below 1 $m\ s^{-1}$ in the short term.

The project was presented to the call for new instrumentation published by CAHA in May 2018, receiving a positive evaluation from a scientific committee. Technical development began in mid-2020, though the implementation of upgrades was constrained by the instrument's scientific operation requirements since early 2016.

The primary focus of C-PLUS was on the NIR channel's CS due to its significant potential for thermal stability improvement. Nevertheless, other subsystems such as the calibration unit (including the Fabry-pérot etalon and the optical fibre injection arrangement), the VIS spectrograph, the coudé room conditioning, and the pipeline will also benefit from C-PLUS.

### 3.1 CARMENES-PLUS upgrades and new thermal stability

To evaluate the stability of the spectrograph after implementing the C-PLUS upgrades, we focused on key physical variables such as pressure and temperature. The analysis distinguishes between pre-C-PLUS (before any upgrades, from 1 November 2016 to 7 February 2021) and post-C-PLUS (after fully upgraded, from 1 June 2022 to 29 June 2024).

Data from strategically placed sensors [26, also Fig. 3 in this work] were used to assess the improvements. The temperature at the entrance of the first flow splitter (NIR-CS-Ts02, hereafter Ts02[2]) monitors the temperature of the coolant entering the cooling lines. The sensor at the first heat exchanger (Ts06) shows the variability of the temperature during the thermal exchange. The sensor located at the middle point of the upper radiation shield (Ts19) provides information from a location with a smooth and averaged thermal gradient, offering good information about thermal drifts. A temperature sensor at the first heat exchanger of the first cooling line of the lower radiation shield (Ts22) is used as reference for the setpoint of the control loop. The temperature of the optical bench (Ts10) is also monitored since it is the ultimate target for thermal stabilisation.

Temperature data from Ts10 and Ts19 is shown in Fig. 4, with a temperature scale selected to highlight details in the analysed periods (before and after C-PLUS). During the C-PLUS interventions, the temperature variations were larger (greyed out in Fig. 4). For pressure analysis, full-range vacuum gauges monitor the vacuum level of the transfer lines, and a pressure gauge (NIR-CS-S1) tracks the feeding pressure in the fixed-dewar.

Data from pre- and post-C-PLUS configurations, as shown in Fig. 5, quantifies the improvement in thermal stability. Two main trends were identified: short-term variations (intra-night or over hours) and long-term variations (over days to weeks).

[2] All sensors will hereafter be identified in the text with the nomenclature Ts (Temperature sensor) and its corresponding number: Ts02.

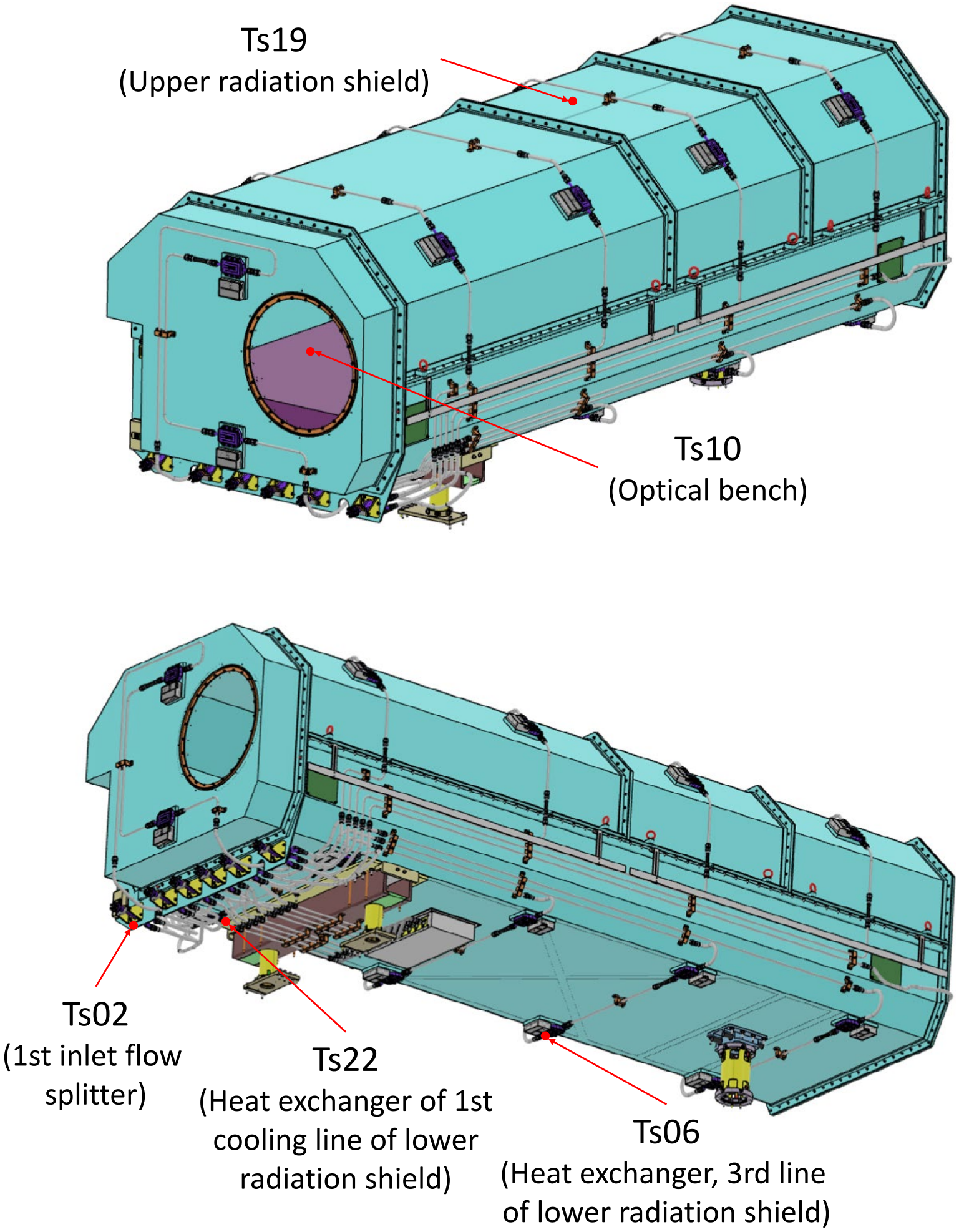

**Fig. 3** Location of key temperature sensors (Ts) in the CARMENES-NIR cryogenic system. The top panel shows the upper radiation shield (Ts19) and the optical bench (Ts10) sensors. The bottom panel displays the underside of the instrument, highlighting sensors placed at the 1st inlet flow splitter (Ts02), and at heat exchangers of the 1st (Ts22) and 3rd (Ts06) cooling lines of the lower radiation shield

In both cases, the amplitude of variation is smaller during the post-C-PLUS period. To ensure comparability between the two configurations, a period without noticeable thermal drifts was selected, excluding transient states caused by tests and interventions conducted during the project.

The following subsections detail the C-PLUS upgrades one by one and their impact on thermal stability.

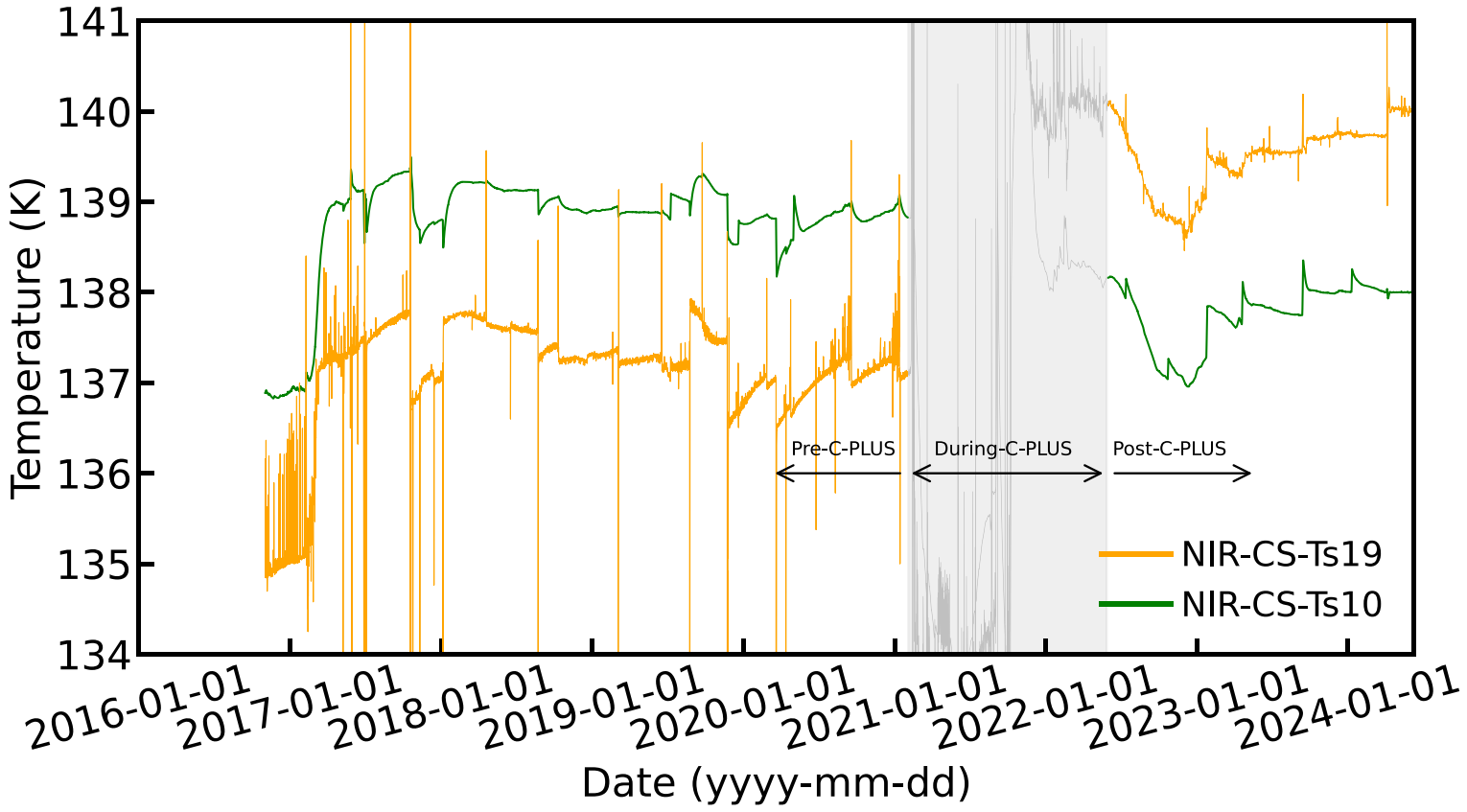

**Fig. 4** Data history of the temperature sensors NIR-CS-Ts10 (at the optical bench, green line) and NIR-CS-Ts19 (at the radiation shield, orange line). The shaded region indicates the period during which the C-PLUS interventions took place, distinguishing between the pre- and post-C-PLUS phases. A noticeable temperature disturbance is observed during the C-PLUS technical campaign due to interventions and transient events

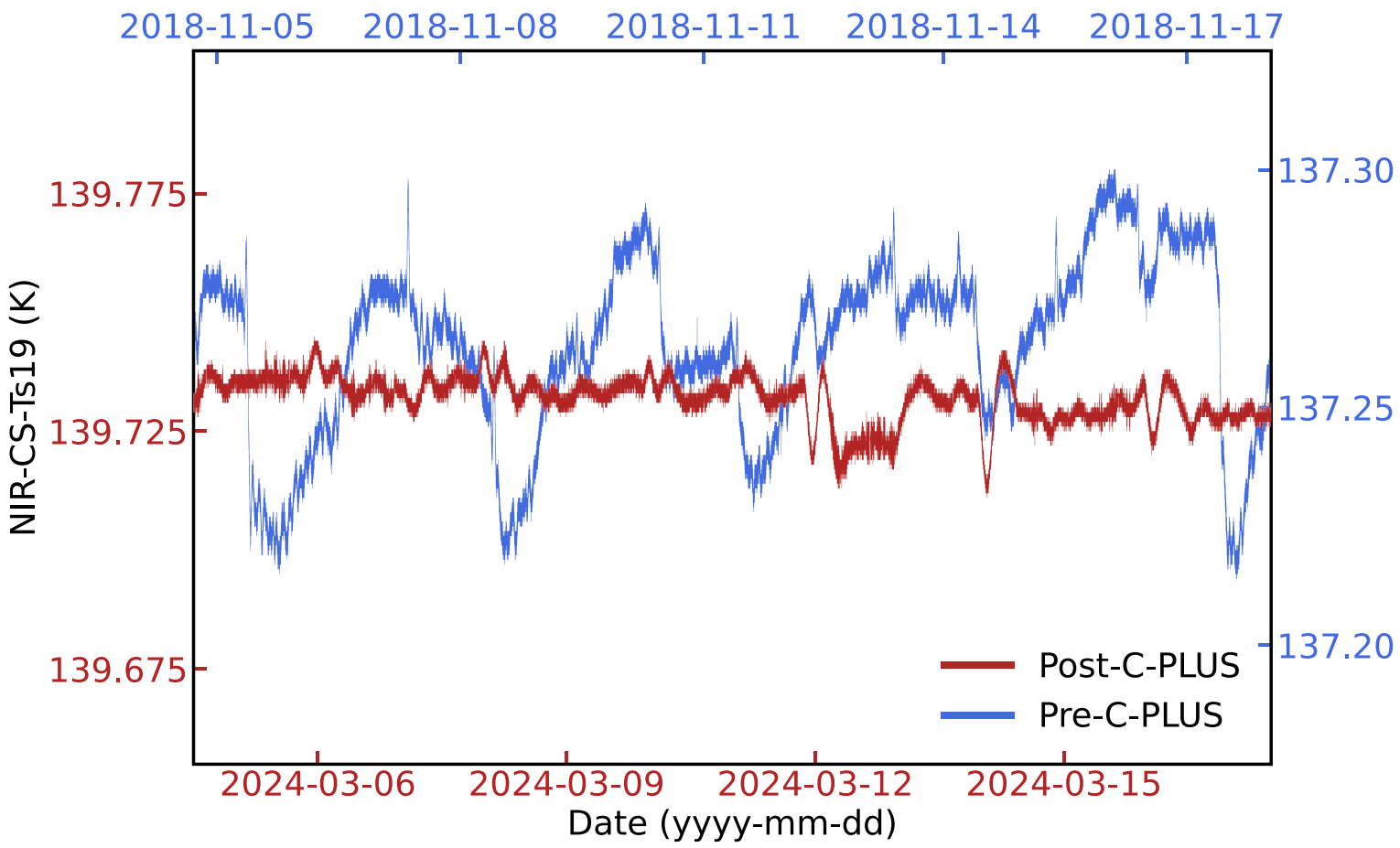

**Fig. 5** Temperature measured at a central point of the radiation shield (NIR-CS-Ts19), shown for the pre-C-PLUS (blue) and post-C-PLUS (red) periods. After the C-PLUS upgrades, both short-term fluctuations and long-term drifts are significantly reduced

### 3.1.1 Automatic vacuum system for transfer lines (AVSTL)

The transfer lines (item 4 in Fig. 1) carrying nitrogen gas below 130 K are vacuum-encased to maintain low temperatures. The internal vacuum, initially preserved by a blind lid, degraded over time due to leakage and out-gassing, negatively affecting thermal performance. This degradation was revealed when frost formed on the lines ends (no vacuum gauge was available). Opening the lid for manual vacuum restoration showed very poor vacuum levels inside the line (pressure values of even 0.1 mbar). These manual restorations back to high vacuum conditions caused abrupt thermal changes.

An automatic vacuum system for transfer lines (AVSTL, Fig. 6) was developed to maintain a high vacuum level, setting an upper-pressure limit requirement of $1 \cdot 10^{-5}$ mbar. However, the good sealing of the AVSTL and the cold trap effect maintain the vacuum level for long periods within the range of $5.9 \cdot 10^{-7}$ to $9.1 \cdot 10^{-7}$ mbar.

This upgrade minimised the thermal drift of the radiation shield caused by gradual vacuum degradation and eliminated thermal instabilities due to manual vacuum regeneration every few months, leading to smoother variations of the coolant temperature.

### 3.1.2 Replacement of the on/off valve by a proportional one

The on/off valve used to control the gas flow discontinuously, causing temperature oscillations at the radiation shield, directly affecting RV measurements. Replacing it with a proportional valve changed the CS from a discontinuous to a continuous flow mode, avoiding coolant flow interruptions. A control loop was implemented to adjust the proportional valve's aperture based on the Ts22 sensor's setpoint, ensuring smoother thermal conditions. The peak-to-peak variations of the temperature decreased from $\Delta T = 0.09$ K to $\Delta T = 0.002$ K. The character of the temperature variations also changes. With the original on/off valve, the signal shows a saw-tooth pattern, while with the new proportional valve, the variations appear as smaller, random fluctuations around a stable set point. Figure 7 demonstrates this with dates chosen to show the typical performance.

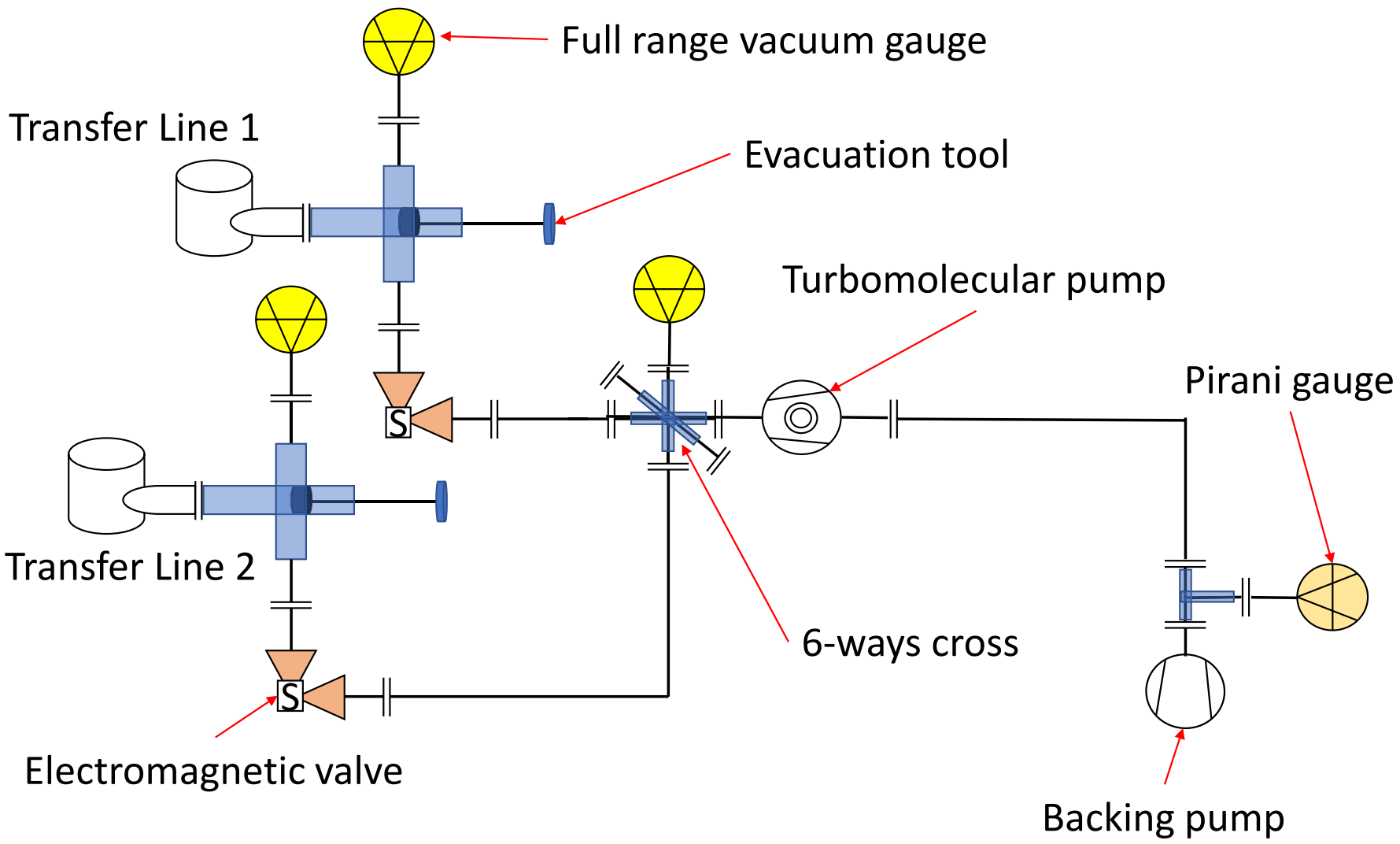

**Fig. 6** Layout of the automatic vacuum system for the transfer lines; it includes a pumping group that generates vacuum, electromagnetic valves that open and close depending on vacuum levels, and vacuum gauges for continuous pressure monitoring. A manual valve (evacuation tool) allows opening the transfer lines without breaking the internal vacuum

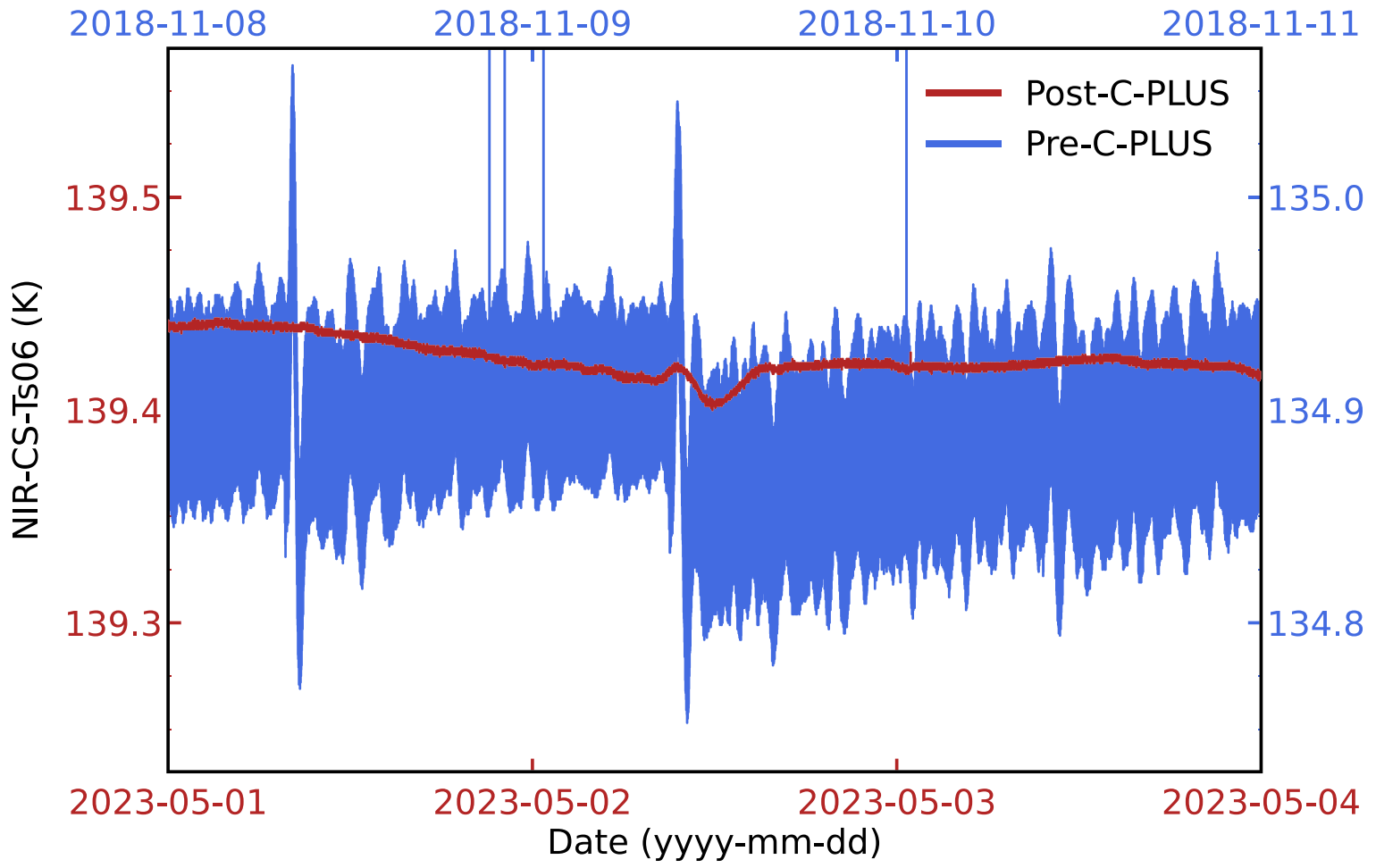

**Fig. 7** Temperature variability over 3-day intervals at the heat exchanger attached to the radiation shield (NIR-CS-Ts06), shown for pre-C-PLUS (blue) and post-C-PLUS (red) periods. After the C-PLUS upgrades, the temperature variations are significantly reduced and display a much smoother profile

An important advantage of having a proportional valve is keeping the pressure gradient constant from the 350-l dewar to the exit valve. This is a considerable issue to be taken into account because the pressure in the 350-l dewar undergoes a pressure drop while the LN2 level goes down, being a source of instabilities in the CS.

Nevertheless, the proportional valve configuration presents a higher sensitivity to pressure variability inside the fixed-dewar during refills in the fixed-dewar configuration. During the refilling process, turbulence encourages evaporation, causing the pressure to increase inside the fixed-dewar and resulting in a slight increase in flow that over-cools the system. The pressure control unit (PCU; see Subsection 3.1.3 below) is a key feature to counteract these pressure fluctuations, mitigating the pressure variability and ensuring stable operation.

### 3.1.3 Pressure control unit (PCU)

The pressure control unit (PCU; sketch in top panel of Fig. 8) uses nitrogen gas from a 150-l dewar pressurised at 0.8 bar. If the 350-l dewar's pressure drops below the lower limit, the PCU injects gas; if the pressure exceeds the upper limit, the PCU evacuates excess gas. These processes are controlled by two pneumatic proportional valves using a proportional-integral-derivative control for fine-tuning.

Several previous tests concluded that pressures below 0.36 bar in the 350-l dewar feeding the N2GPU led to unstable operation. The PCU stabilises the coolant flow and the pressure inside the dewar. As shown in bottom panel of Fig. 8, the pressure is maintained within the peak-to-peak range of 0.421 to 0.426 bar, compared to the previous range of 0.370 to 0.437 bar before its implementation.

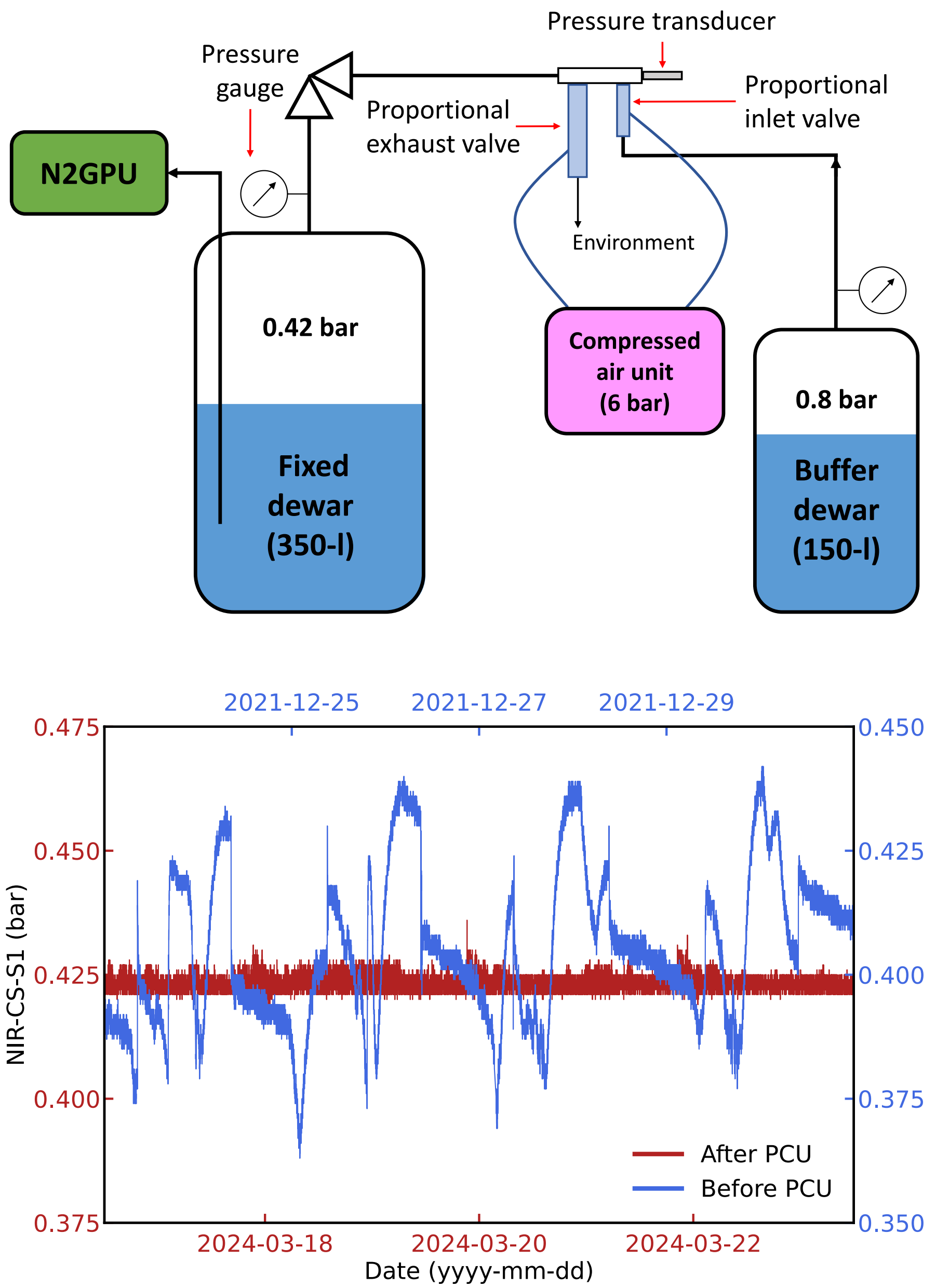

**Fig. 8** *Top*: Schematic layout of the main components of the Pressure Control Unit (PCU) in the CARMENES-NIR cooling system. A proportional valve regulates the gas flow from the buffer dewar to the fixed one to maintain the latter at a stable pressure. Pressure is continuously monitored using pressure transducers and gauges. *Bottom*: Pressure inside the fixed dewar over a representative 7-day period, shown before (blue) and after (red) the implementation of the PCU. The post-upgrade data demonstrate significantly reduced fluctuations and improved stability around the setpoint

### 3.1.4 Fixed-dewar configuration

The daily replacement of the dewar feeding the N2GPU significantly impacted the CS stability due to interruptions in LN2 flow, causing sudden changes in thermal stability and resulting in RV drifts. Additionally, different working pressures in the various dewars used caused daily pressure jumps, leading to intra-day thermal drifts. To allow these instabilities to settle before night observations, dewar replacements were performed early in the morning.

To minimise these disruptions, a fixed-dewar feeding the N2GPU continuously, along with a supporting removable dewar feeding the fixed-dewar, was proposed. In this configuration (top panel in Fig. 9), the flow from the supporting dewar is controlled by the pressure difference between the two LN2 dewars, with the supporting dewar maintained at 0.95 ± 0.05 bar. The feed line transfers LN2 from the supporting to the fixed-dewar at a rate of 80 kg $h^{-1}$ under vacuum conditions.

A modified manifold allows the fixed-dewar to feed and be refilled simultaneously through two independent circuits. The refilling circuit has two pipes to prevent narrowing and maintain a section equal to or greater than the feed line from the supporting dewar. The LN2 is injected horizontally at the bottom of the dewar through both pipes in opposite directions, and 2 cm above the N2GPU feeding pipe, ensuring no interaction between the processes. The LN2 flows from the bottom of the fixed-dewar to the N2GPU through a central pipe.

The temperature sensor Ts02 at the inlet flow splitter allows comparison of the impact of both dewar configurations on the coolant temperature, as illustrated in the bottom panel of Fig. 9. We selected periods of 10 days with representative performance in both cases. The temperature perturbation due to refills in the fixed-dewar configuration is significantly less abrupt than that of dewar replacements, with temperature fluctuations reduced from $\pm 3$ K during dewar replacements to $\pm 0.4$ K during fixed-dewar refills. To further enhance thermal stability, the interval between refills was extended from one day to two days in November 2022, as the dewar capacity was sufficient to maintain this longer interval.

### 3.1.5 Feed line's check valve removal

The N2GPU is fed through an in-vacuum feed line. To ensure LN2 flow from the fixed-dewar only when the feed line is connected, a safety mechanism with two check valves in the Johnston coupling (top panel of Fig. 10) is used. This type of connection is widely employed in vacuum-insulated piping systems for cryogenic fluids. One valve is located in the inner part of the dewar's manifold (female port), and the other is in the feed line's lance (male port). When the lance is plugged into the manifold, both check valves open, allowing LN2 to flow.

During operation, stochastic interruptions of LN2 flow caused significant thermal instabilities. These interruptions originated from pressure drops due to excessive narrowing of the check valve on the feed line side (male) compared to the manifold side (female; see bottom panel of Fig. 10). To eliminate this bottleneck, the check valve on the feed line side was removed and replaced with a fixed sleeve bushing, which is geometrically identical to the check valve.

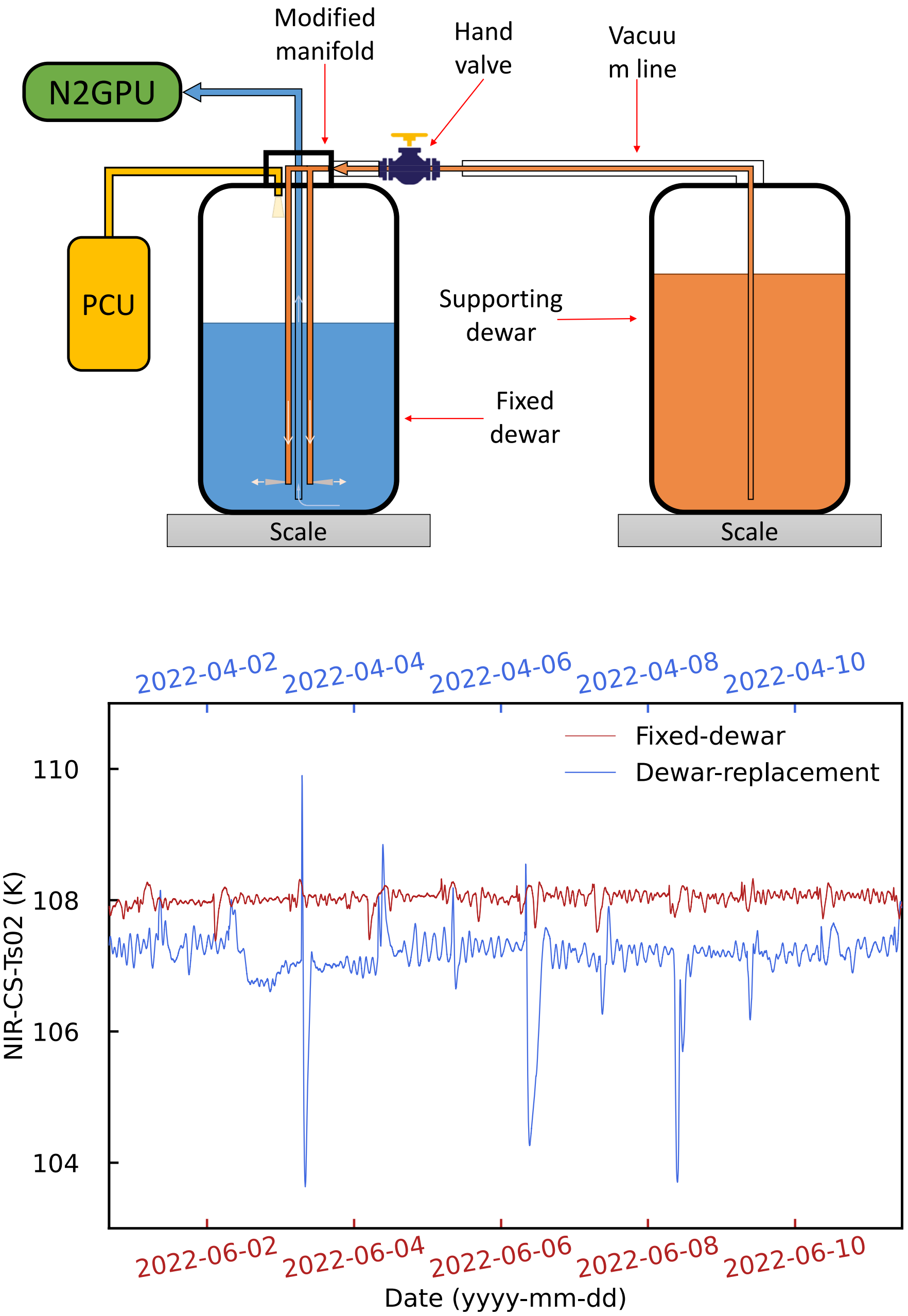

**Fig. 9** *Top*: Layout of the system in the fixed-dewar configuration. The supporting dewar is shown on the right, while the fixed dewar on the left feeds the N2GPU and includes the pressure control unit. *Bottom*: Coolant temperature at the inlet flow splitter over two representative 10-day periods, comparing the dewar-replacement configuration (blue) and the fixed-dewar configuration (red). The new setup significantly reduces the amplitude of temperature fluctuations and minimizes abrupt changes during refilling or replacement operations

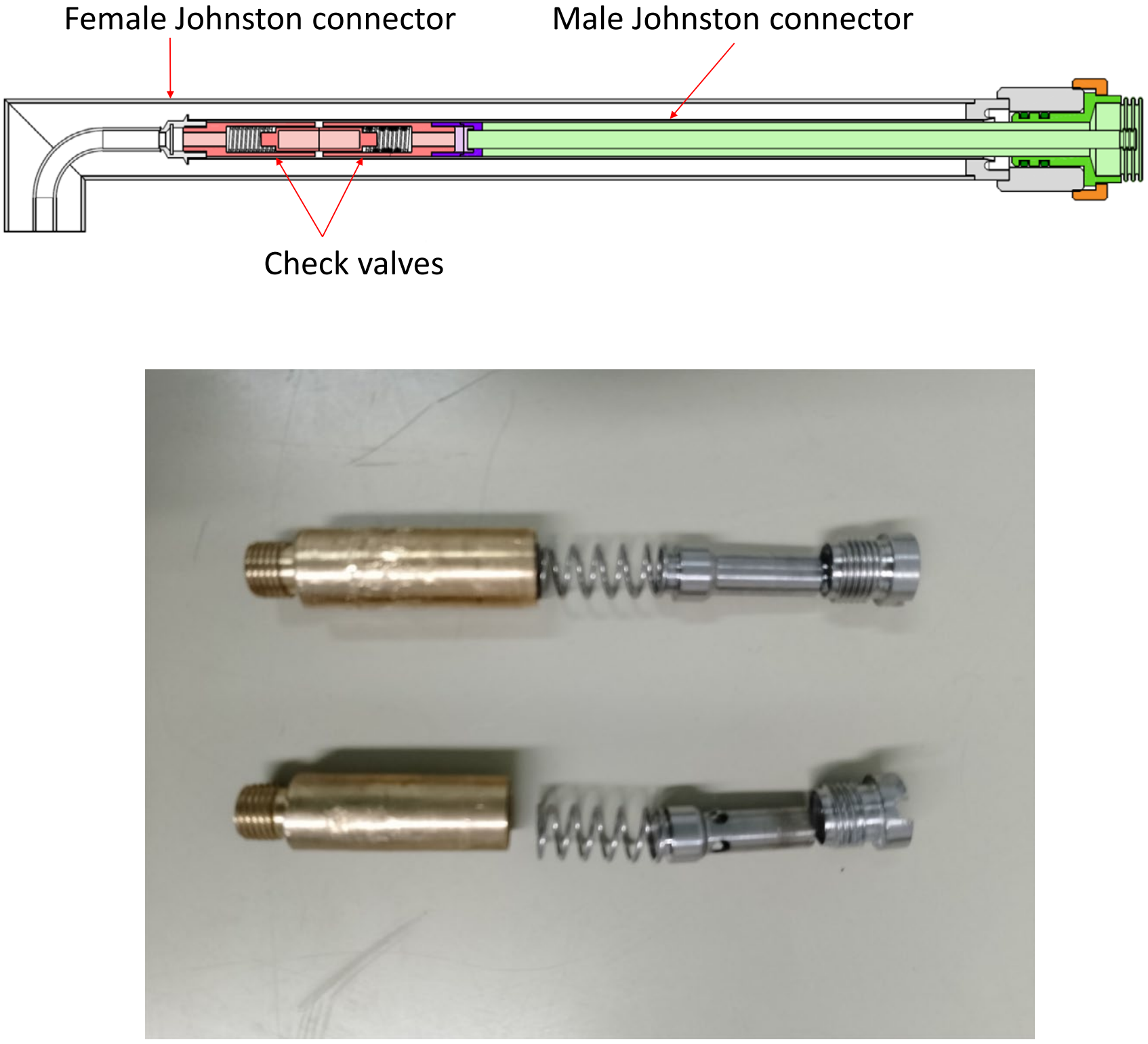

**Fig. 10** *Top*: Layout of the Johnston coupling. Note the check valves highlighted in pink. *Bottom*: Photograph of the feed line's check valve (top) and the manifold's check valve with orifices (bottom)

The removal of the feed line's check valve eliminated the narrowing in the circuit and the random interruptions of LN2 flow. A side effect was a smooth thermal drift of the instrument over several months (at a rate of 0.015 K day$^{-1}$, within the requirements; [27]) towards a new steady state. Once the new steady state was reached, the radiation shield temperature variations with the continuous flow configuration were considerably smoother and smaller than those with the discontinuous configuration (see Fig. 5).

## 4 Data reduction and calibration

In this section, we describe the data reduction process and introduce the datasets used to assess the impact of the C-PLUS thermal upgrades. Our analysis is based on three types of data: (i) Fabry-Pérot calibration spectra, which allow us to quantify the intrinsic instrumental stability; (ii) spectra of RV-constant stars (rms $\leq$ 10 m s$^{-1}$), used to evaluate the night-to-night repeatability of the RV measurements; and (iii) on-sky stellar observations, which reveal the overall improvement in RV precision for scientific targets. These datasets provide a comprehensive picture of the improvements achieved by C-PLUS, both from an engineering and a scientific perspective.

### 4.1 Data reduction

The CARMENES spectra are reduced using the caracal (CARMENES Reduction And CALibration) pipeline, as detailed in [28]. Starting with the raw data, caracal performs dark/bias correction, order tracing, flat-relative optimal extraction, and wavelength calibration, resulting in fully reduced and wavelength-calibrated 1D spectra.

The second pipeline is serval (SpEctrum Radial Velocity AnaLyser), described in detail by [29]. It creates a high signal-to-noise ratio template spectrum using the caracal reduced spectra of the target star. It then computes the series of RVs via least-square fitting, telluric masking, échelle order weighting and correcting for systematic errors. Apart from the RVs, the output also includes activity indices.

### 4.2 Calibration strategy and process

Accurate wavelength calibration and precise monitoring of instrument stability are crucial for obtaining high-precision RV measurements. Each of the two CARMENES spectrographs use two fibres (A and B) to feed star, sky, and calibration light beams into each spectrograph. These two beams are then focused onto the detectors, providing two adjacent spectra that can be calibrated against each other [7].

The baseline strategy for observation involves injecting starlight into fibre A and light from a calibration source into fibre B, enabling simultaneous calibration. This setup allows continuous monitoring and correction of any instrumental drifts during observations. By calibrating the spectra against each other, wavelength shifts can be accurately tracked and corrected.

A study by [30] [see also 31] demonstrates that using a combination of hollow cathode lamps (HCLs) and a Fabry-Pérot interferometer (FP) is more effective for wavelength calibration of échelle spectrographs than using HCLs alone. The FP produces a denser grid of lines with nearly uniform intensity across the spectral range, enhancing calibration accuracy. Consequently, CARMENES adopted this technique [32].

At the beginning and end of each night, HCL images are acquired to derive the wavelength solution at the detector. This solution is then used to calibrate the FP spectra, assigning wavelengths to each interferometry peak. The FP spectra are used to monitor the drift of the wavelength solution due to thermo-mechanical changes within the instrument during the night.

The nightly RV drift is defined by the data obtained from each FP calibration throughout a full CARMENES observing night. Nights with calibrations only at the start and end are excluded from this analysis. Figure 11 shows the complete history of the nightly FP RV calibrations collected.

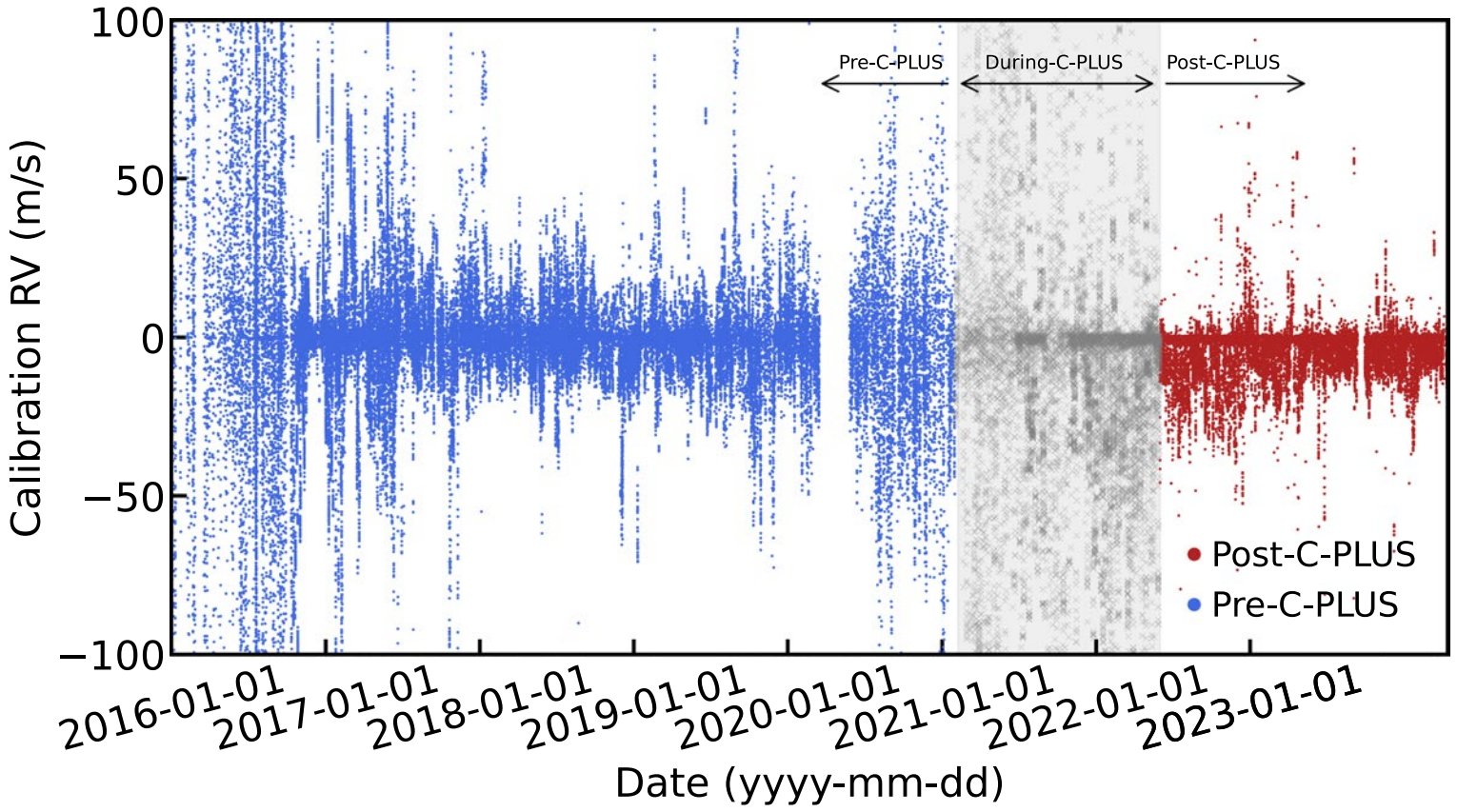

**Fig. 11** Data history of nightly FP RV calibrations for the CARMENES NIR channel. The three operational periods are marked in blue (pre-C-PLUS), grey (during-C-PLUS), and red (post-C-PLUS). Variations peak during the C-PLUS intervention period and are notably reduced afterward, both in amplitude and in the number of outliers, compared to the pre-C-PLUS phase. Complete RV calibration datasets for both the NIR and VIS channels are available online

### 4.3 Selection of on-sky datasets

The RV time series from the 361 M dwarf stars observed by CARMENES during its guaranteed time observations (GTO) phase [2016-2020; 8] are publicly available[3]. CARMENES has primarily targeted early- and mid-type M dwarfs with near-solar metallicity [19, 33, 34].

For the analysis presented here, we used the serval rvc data-product, which are RVs corrected for the instrument's drift and secular acceleration, but not for nightly zero points (NZPs; see Sect. 4.4). We limit our analysis to slowly rotating M dwarfs ($v \sin i_\star \leq 2$ km s$^{-1}$) with at least 5 spectra. These datasets can have from 5 to hundreds of observations, depending on how often a target has been visited (see Table 1). The $v \sin i$ values are sourced from [35], were an upper limit of 2 km s$^{-1}$ for $v \sin i$ is set, as values below this threshold cannot be reliably obtained [19]. These criteria reduce the stellar sample size to 174 stars for pre-C-PLUS and 38 stars for post-C-PLUS. For consistency, we used the same stellar sample for the VIS channel as for the post-C-PLUS NIR sample (174 stars). This selection defines the three datasets of on-sky spectra used in this work.

### 4.4 Nightly zero points

Despite the CARMENES spectrograph being wavelength-calibrated each afternoon and its drift monitored using the FP etalon during the night (see Sect. 4.2), on-sky RV measurements from a given night still share common systematic offsets, known as nightly zero points (NZPs) [8]. These offsets arise from a combination of FP drifts, degradation in the quality of the hollow cathode lamp (HCL) reference spectra, instru-

[3] http://carmenes.cab.inta-csic.es/gto/jsp/dr1Public.jsp

**Table 1** Stellar RV jitter estimates and number of measurements for a subset of CARMENES targets

| Karmn | $\sigma_{vis}$ | $N_{obs}$ | $\sigma_{nir-pre}$ | $N_{obs}$ | $\sigma_{nir-post}$ | $N_{obs}$ |
|---|---|---|---|---|---|---|
| | [ m s$^{-1}$] | VIS | [ m s$^{-1}$] | NIR-pre | [ m s$^{-1}$] | NIR-post |
| J17033+514 | 4.3 ± 0.4 | 65 | 4.3 ± 2.9 | 34 | 6.7 ± 1.1 | 31 |
| J22565+165 | 3.3 ± 0.1 | 715 | 4.5 ± 0.5 | 442 | 2.8 ± 1.0 | 180 |
| J06548+332 | 3.0 ± 0.1 | 465 | 3.0 ± 0.9 | 261 | 2.1 ± 1.3 | 122 |
| J05033-173 | 4.0 ± 0.4 | 76 | 6.0 ± 1.5 | 46 | 3.4 ± 2.0 | 27 |
| J02530+168 | 2.3 ± 0.1 | 301 | 3.2 ± 0.4 | 255 | 3.3 ± 0.8 | 61 |
| J23492+024 | 1.5 ± 0.1 | 493 | 3.6 ± 1.0 | 270 | 3.3 ± 1.1 | 136 |
| J13299+102 | 2.0 ± 0.1 | 458 | 4.4 ± 1.7 | 299 | 2.7 ± 1.3 | 89 |

For each star (Karmn ID), we list the RV jitter (in m s$^{-1}$) and number of observations in the VIS channel, and the NIR channel before and after the C-PLUS upgrades. The table contains seven stars that have jitter estimate for all three cases; the full list is available online ment instabilities during calibration (due to the FP and HCL not being usable simultaneously), and variations in calibration light injection caused by imperfect scrambling.

We calculate the NZPs using a sample of RV-constant stars, defined as those with an RV time series scatter of rms $\leq$ 10 m s$^{-1}$, following the procedure of [36]. This method involves selecting stars with minimal RV variation over time as reference points, computing the nightly offset required to align these stars' RV measurements with their known values, and applying this offset to correct the RV measurements of all stars observed that night. To prevent self-biasing, a star's own RV measurements are excluded when computing the NZP for its night. If fewer than three RV measurements of RV-constant stars were obtained on a given night, the NZP is replaced by the median of all NZP values over the survey period [8].

The computed NZP offsets are then subtracted from the serval RV measurements (rvc) to produce the final NZP-corrected RV values, ready for scientific analysis.

## 5 Results and discussion

In this section, we present the results of our analysis, quantifying the impact of the C-PLUS upgrades on the RV performance of the NIR channel. We begin in Sect.5.1 discussing the intrinsic RV precision derived from calibration data (as described in Sect.4.2), which reflects the instrument's thermal and mechanical stability. Next, in Sect.5.2, we examine the scatter in the NZPs, using spectra from RV-constant stars to probe night-to-night stability (see Sect.4.4). Finally, in Sect.5.3, we analyse the RV scatter in stellar observations (see Sect.4.3), highlighting the improvement in on-sky performance and scientific precision after the upgrades.

### 5.1 Intrinsic RV precision

In the work of [23], an analysis of the instrumental performance of both CARMENES spectrographs based on calibration data was carried out. Their conclusion was that, at that time, the VIS channel achieved approximately three times better intrinsic precision than the NIR, with values of 1.2 m s$^{-1}$ and 3.7 m s$^{-1}$, respectively.

This subsection focuses on the improvement in the intrinsic RV precision of the CARMENES NIR spectrograph following the implementation of the C-PLUS upgrades. The precision is quantified by the scatter of the nightly RV drift obtained from FP calibration spectra (see Sect. 4.2).

Figure 12 shows the evolution of these calibration RVs, with a vertical line indicating the date when the fixed-dewar configuration was introduced. The plot includes data from 3.5 months before and after this change, clearly illustrating the reduction in the amplitude of the nightly RV drifts resulting from the improved thermal stability provided by the fixed-dewar setup.

To quantify the improvement in RV stability, we considered two statistical properties of the nightly RV drifts: the peak-to-peak amplitude and the root-mean-square of the residuals ($rms_{res}$) after fitting a quadratic least-squares model to the data.

The peak-to-peak value reflects the total RV variation induced by the instrument's drift during a single night, while the $rms_{res}$ quantifies the scatter of the measurements around the fitted trend. Throughout this work, we adopt the $rms_{res}$ as a measure of the intrinsic RV precision, since it represents the residual uncertainty after applying the best correction for the instrumental drift. The smoother the nightly drift and the lower its amplitude, the more precisely the FP calibration can correct the measured stellar RVs.

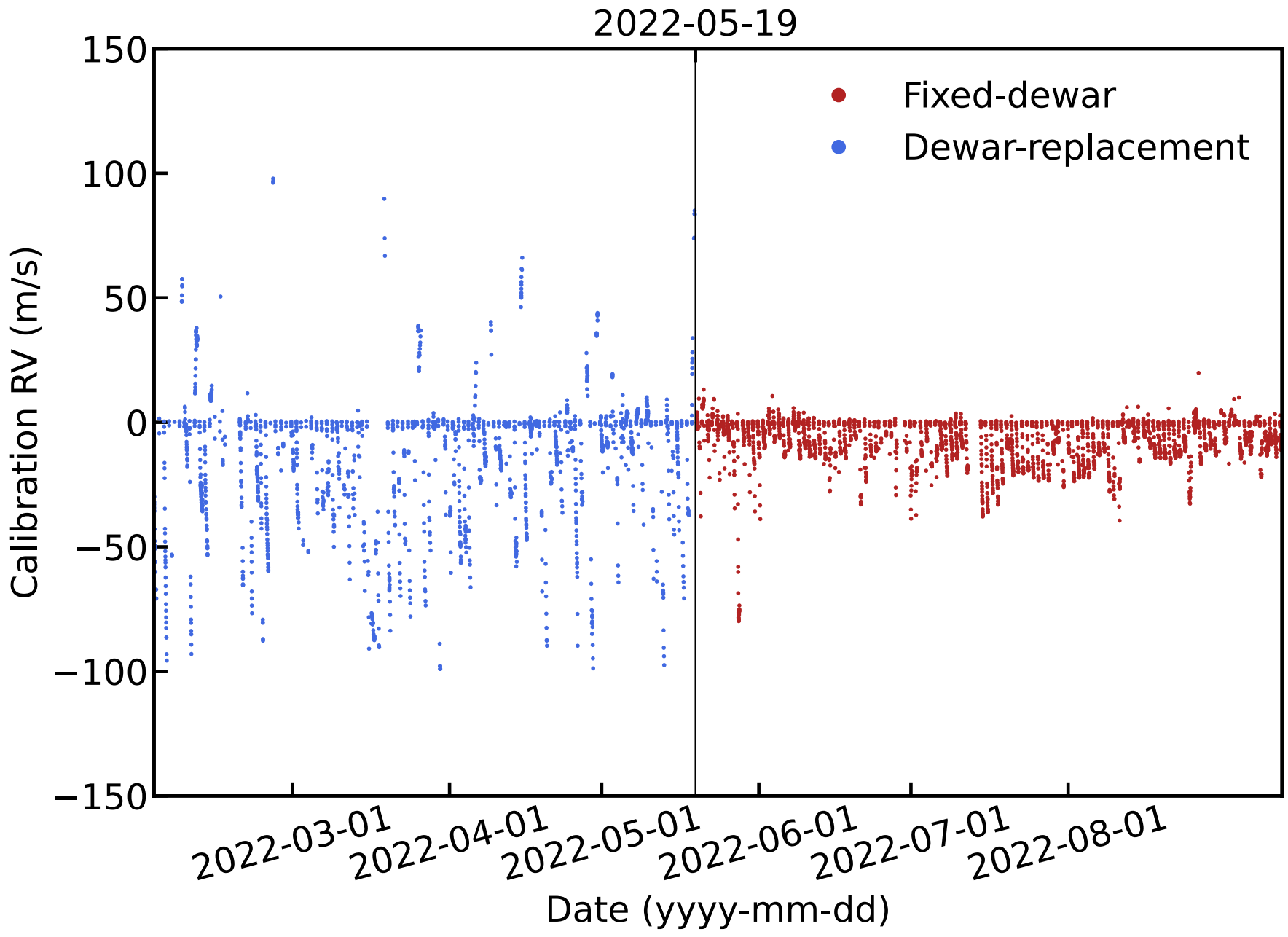

**Fig. 12** RV measurements derived from Fabry-Pérot calibration spectra, illustrating the impact of the C-PLUS fixed-dewar configuration on the thermal stability of the NIR spectrograph. The vertical line indicates the implementation date of the fixed-dewar system. Data from 3.5 months before (blue, dewar-replacement configuration) and 3.5 months after (red, fixed-dewar configuration) are shown. Note the reduced intra-night and night-to-night RV variations following the upgrade

Figure 13 illustrates the quadratic fitting for a series of three consecutive typical nights, with the top panel representing pre-C-PLUS and the bottom panel post-C-PLUS. It is clear that the amplitude and scatter of the data have decreased after C-PLUS. Additionally, the drifts are now more consistent and similar from one night to another.

We performed a statistical analysis of both the peak-to-peak and $rms_{res}$ values derived from the nightly RV drifts. We analyse three datasets: CARMENES NIR pre- and post-C-PLUS (see Sect. 3) and CARMENES VIS (from 1 November 2016 to 29 June 2024). Their histograms are represented in Fig. 14.

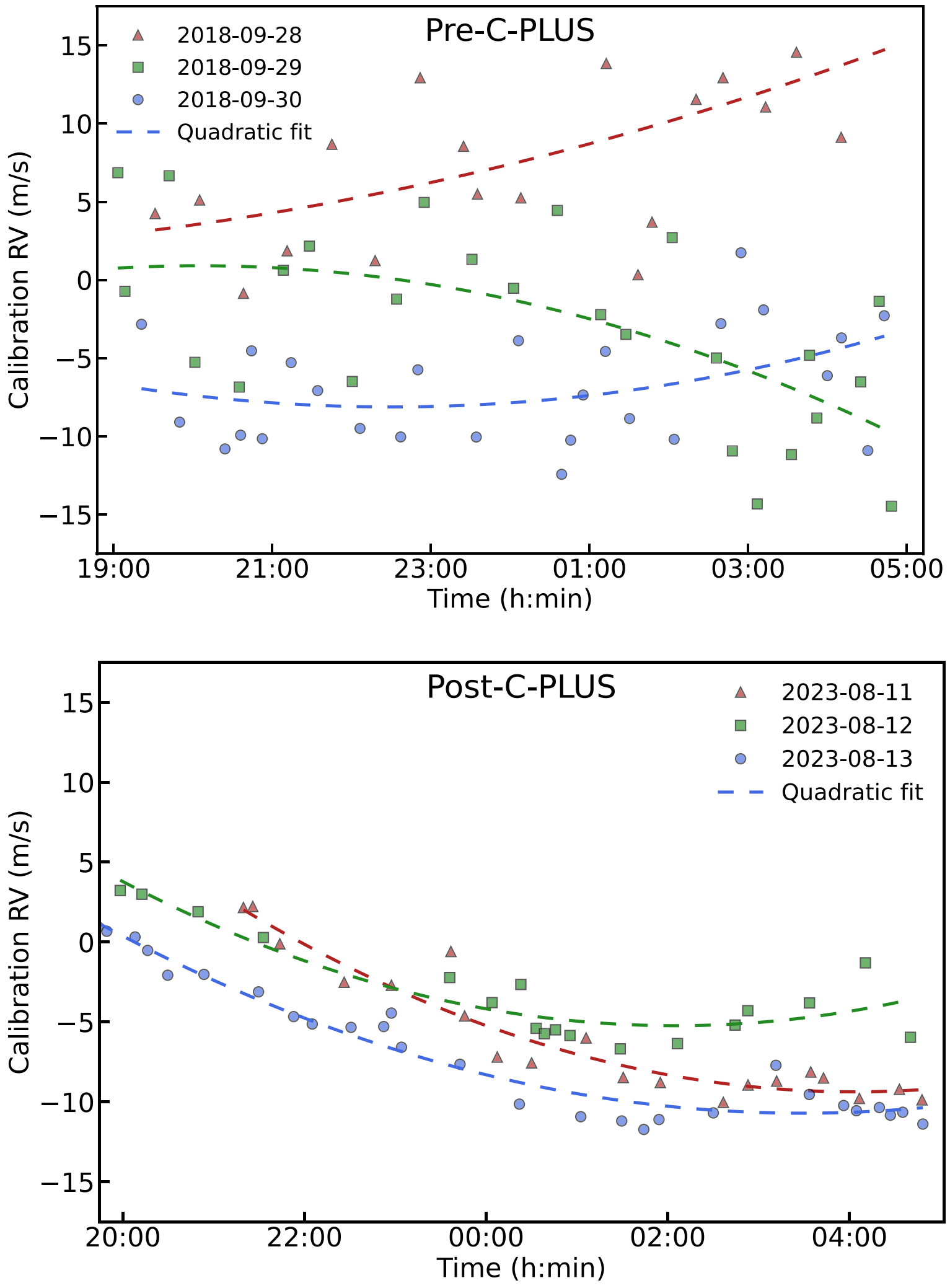

**Fig. 13** RV drift of the NIR spectrograph during individual nights, derived from Fabry-Pérot calibration spectra. The panels show three consecutive nights of pre-C-PLUS (top) and post-C-PLUS (bottom) to illustrate the typical drift before and after the upgrades. The dashed lines represent the quadratic fits applied to the calibration data for each night

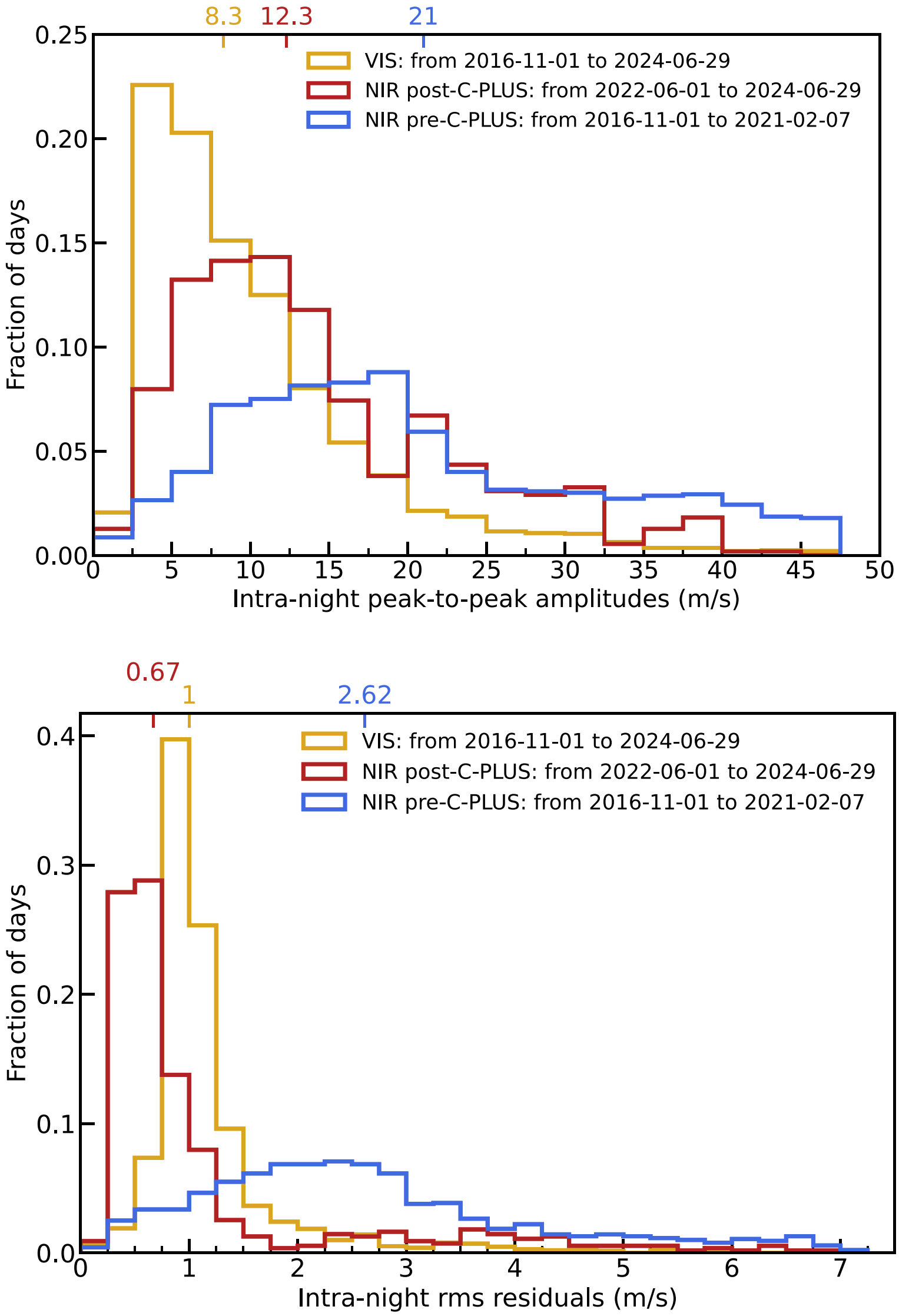

**Fig. 14** *Top*: Histogram of the peak-to-peak RV variation from Fabry-Pérot calibrations. *Bottom*: Histogram of the root-mean-square of the residuals after subtracting a quadratic fit. Results are shown for CARMENES VIS (yellow), NIR pre-C-PLUS (blue), and NIR post-C-PLUS (red). The vertical axis indicates the fraction of days (normalized to 1), and the median values are indicated on the top axis. Note the significant reduction in the median and tighter distribution after the C-PLUS upgrades

The top panel of Fig. 14 displays the histograms of the peak-to-peak RV amplitudes. All three distributions are unimodal, indicating that the data do not exhibit significant substructures or clustering around multiple distinct values. The pre-C-PLUS distribution spans a wide range, with a median value of 21.0 m s$^{-1}$. It is skewed towards higher amplitudes, suggesting that a non-negligible fraction of nights were affected by unusually large RV drifts. The variance is correspondingly high, reflecting the broad spread around the median. In contrast, the post-C-PLUS distribution shows a substantially narrower spread, with a reduced median value of approximately 12.3 m s$^{-1}$, indicating improved intra-night stability following the instrument upgrades. The VIS channel displays the lowest median drift amplitude at 8.3 m s$^{-1}$, and its distribution is the most compact among the three.

The lower panel of Fig. 14 presents the distributions of the rms$_{res}$ values. These show a similar pattern to the peak-to-peak amplitudes, although the post-C-PLUS distribution is more sharply peaked around lower values compared to both the pre-C-PLUS and VIS datasets. The median rms$_{res}$ values are 2.62 m s$^{-1}$ for the pre-C-PLUS sample, 0.67 m s$^{-1}$ for the post-C-PLUS sample, and 1.0 m s$^{-1}$ for the VIS channel.

These results confirm that the C-PLUS upgrades significantly improved the NIR spectrograph's short-term stability, bringing its performance closer to that of the VIS channel, even better in terms of rms$_{res}$. The more compact and symmetric post-C-PLUS histogram, centred at a lower rms$_{res}$, highlights the enhanced instrument stability and reduced intra-night RV scatter achieved after the upgrade.

## 5.2 Nightly zero point scatter

In this subsection, we quantify the effect of the C-PLUS upgrades on the nightly zero points (NZPs) of the CARMENES NIR channel. Figure 15 presents the time series of NZPs, with the period of C-PLUS implementation and stabilization shaded in grey to distinguish the pre- and post-C-PLUS regimes. To ensure the reliability of the analysis, we exclude NZPs based on fewer than three RV-constant stars, those identified as 3$\sigma$ outliers, and nights without any NZP determination due to the absence of standard stars. These exclusions are necessary, as such values are replaced by the median (see Sect. 4.4), which would otherwise artificially decrease the NZPs scatter.

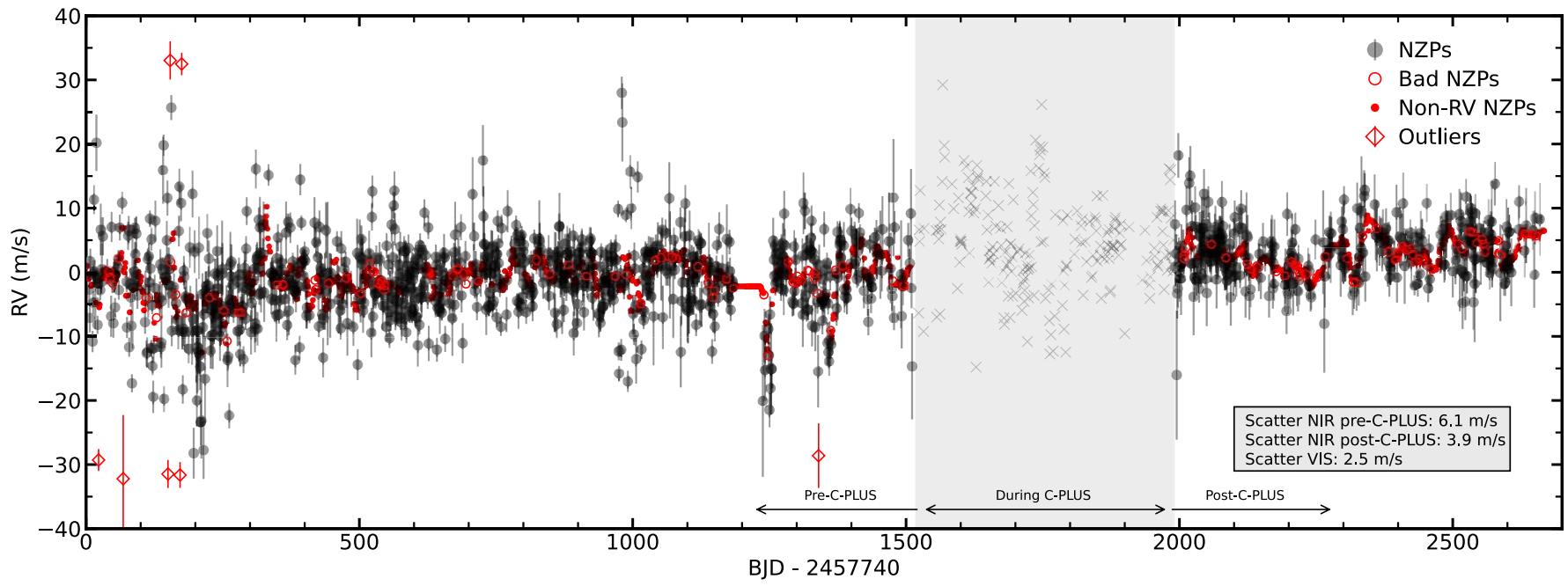

**Fig. 15** Time series of nightly zero points (NZPs) for the CARMENES NIR channel, separated into three intervals; pre-C-PLUS, during C-PLUS, and post-C-PLUS. Solid black circles mark NZPs included in our analysis; crosses within the grey-shaded area indicate values from the C-PLUS intervention period. Red diamonds highlight 3-$\sigma$ outliers; empty red circles denote nights with fewer than three RV-standard observations and solid red dots mark nights without any RV measurements (both replaced by the median NZP). For comparison, the VIS channel NZP scatter value is also shown. Note the reduction in scatter following the C-PLUS upgrades

We computed the NZPs scatter using a maximum likelihood optimisation method developed by [29], mlrms[4]. It accounts for individual measurement uncertainties and allows for an additional unknown variance term, often interpreted as extra jitter or unmodelled noise. This method iteratively adjusts the unknown variance to maximize the likelihood under the assumption of normally distributed residuals.

We find a significant improvement in NZP stability after the upgrades: the scatter decreased from 6.1 m $s^{-1}$ (pre-C-PLUS) to 3.9 m $s^{-1}$ (post-C-PLUS), a reduction of 2.2 m $s^{-1}$. The corresponding median NZP values are $-1.2$ m $s^{-1}$ and 2.9 m $s^{-1}$, respectively. The rate of nights with $3\sigma$ NZP outliers also dropped from 1% pre-C-PLUS to zero post-upgrade. Notably, the average errors of the NZPs are 2.6 m $s^{-1}$ (pre-C-PLUS) and 3.2 m $s^{-1}$ (post-C-PLUS), with the post-C-PLUS scatter closely matching the average error, suggesting that the residual variability may be largely explained by the measurement uncertainties alone.

For comparison, the NZPs in the VIS channel exhibit even lower variability, with a scatter of 2.5 m $s^{-1}$, a median of 0.3 m $s^{-1}$, an outlier fraction of 1.6%, and an average error of 1.1 m $s^{-1}$.

To contextualize these results, we compare them with values reported in the literature for other high-precision spectrographs using similar NZP correction strategies. [37] applied an equivalent approach to HARPS, finding a weighted rms (wrms) scatter of 1.5 m $s^{-1}$. [8] reported a standard deviation of 2.3 m $s^{-1}$ for CARMENES-VIS, in close agreement with our findings. A different approach was taken by [38], who used Gaussian Process regression informed by ancillary parameters to model systematics in SOPHIE's NZP time series. Their method, based on a small set of bright RV-constant stars, yielded an rms of 2.15 m $s^{-1}$.

### 5.3 Contributions to the stellar RV scatter

To understand the sources of RV variability in our on-sky data, we decompose the observed RV scatter into four main contributors: the instrument's intrinsic RV precision, the scatter in the NZPs, the photon noise, and the stellar RV jitter. We then compare the quadratically combined contribution of these components with the median RV scatter measured across the stellar samples defined in Sect. 4.3.

Our approach differs from that of [23] in two important ways. They use RVs corrected for NZP offsets for their analysis, whereas we use uncorrected values, making the NZP contribution explicitly measurable in our study. Additionally, their treatment of stellar jitter assumes a representative value for M dwarfs activity, while we derive the stellar RV jitter individually for each star in our sample, based on its actual time series.

The contribution of the instrument to the RV scatter is here characterised by the $rms_{res}$ of the intra-night drift, which captures short time-scale variations. When the drift is smooth, it can be effectively corrected using simultaneous FP calibration data, making $rms_{res}$ a good proxy for the instrument's intrinsic RV precision.

As described in Sect. 5.2, we estimate the scatter of the NZPs using the mlrms approach, applied after removing clear outliers and poorly estimated values.

[4] https://github.com/mzechmeister/python/blob/master/wstat.py

The photon noise sets a fundamental limit on the RV precision due to the finite number of photons collected during an observation [39]. For our analysis, the photon noise is estimated by serval, based on the signal-to-noise ratio (S/N) of the spectra. The pipeline estimates the RV uncertainty by considering the quality of each spectral order and the overall spectral information content. This provides a robust estimate of the RV uncertainty due to photon statistics. For each star in the sample, we compute the average photon noise across all observations, and then adopt the median of all the targets as the representative value for the sample.

The stellar RV jitter refers to the excess noise in an RV time series caused by apparent Doppler shifts induced by variability in the stellar spectrum [40, 41]. In the work of [42], the RV jitter was computed for most CARMENES VIS targets by modelling it as an additional noise term in the RV error budget. The total uncertainty for each RV measurement was treated as the quadratic sum of its internal error and the stellar jitter. A maximum-likelihood approach was used to simultaneously estimate the jitter and the mean RV, assuming normally distributed residuals. Uncertainties on the jitter were derived via non-parametric bootstrapping with 1000 resamples.

We extended this method to the NIR channel, using RVs corrected for NZPs to ensure that NZP-related variability does not bias the jitter estimate. For each star, the jitter was estimated individually, and we use the median value across the sample distribution as the representative value in our analysis.

Looking at the individual stellar RV jitter values (see Table 1), we find several stars for which the estimated stellar jitter is negative. This indicates that the observed scatter in the RV time series is smaller than expected from the assigned internal uncertainties. The internal errors are primarily driven by photon noise, with additional contributions from the instrument's stability and the data reduction process. A negative stellar jitter estimate typically suggests that the internal errors were overestimated, or that the sample size is too small to reliably constrain the true external scatter. Because a star-by-star analysis is beyond the scope of this work, and the median stellar jitter is not a primary focus here, we assume that potential underestimations of the stellar jitter are balanced by conservatively estimated internal errors.

Based on the four contributors described above, we compute the combined RV scatter as their quadratically summed contribution. The goal of this analysis is not to model the physical origin of each component in detail, but rather to evaluate their relative importance and how they compare to the observed RV scatter in each dataset. To assess the accuracy of this decomposition, we also calculate the median RV scatter of the stellar samples and compare it to the combined value.

We compute the mlrms for the RV time series of the selected stellar samples (see Sect.4.3), and present the resulting distributions in Fig. 16. The quadratically combined RV scatter for the pre-C-PLUS sample is 9.6 m s$^{-1}$, compared to a median measured value of 8.8 m s$^{-1}$. For the post-C-PLUS sample, the combined and median values are 6.9 m s$^{-1}$ and 6.7 m s$^{-1}$, respectively. The CARMENES VIS channel has a combined value of 4.0 m s$^{-1}$ and a median of 4.3 m s$^{-1}$. The discrepancy in pre-C-PLUS data can likely be attributed to significant thermo-mechanical changes in the NIR spectrograph. This might introduce additional instrument jitter, which can affect all contributors and lead to an overestimation of the combined value.

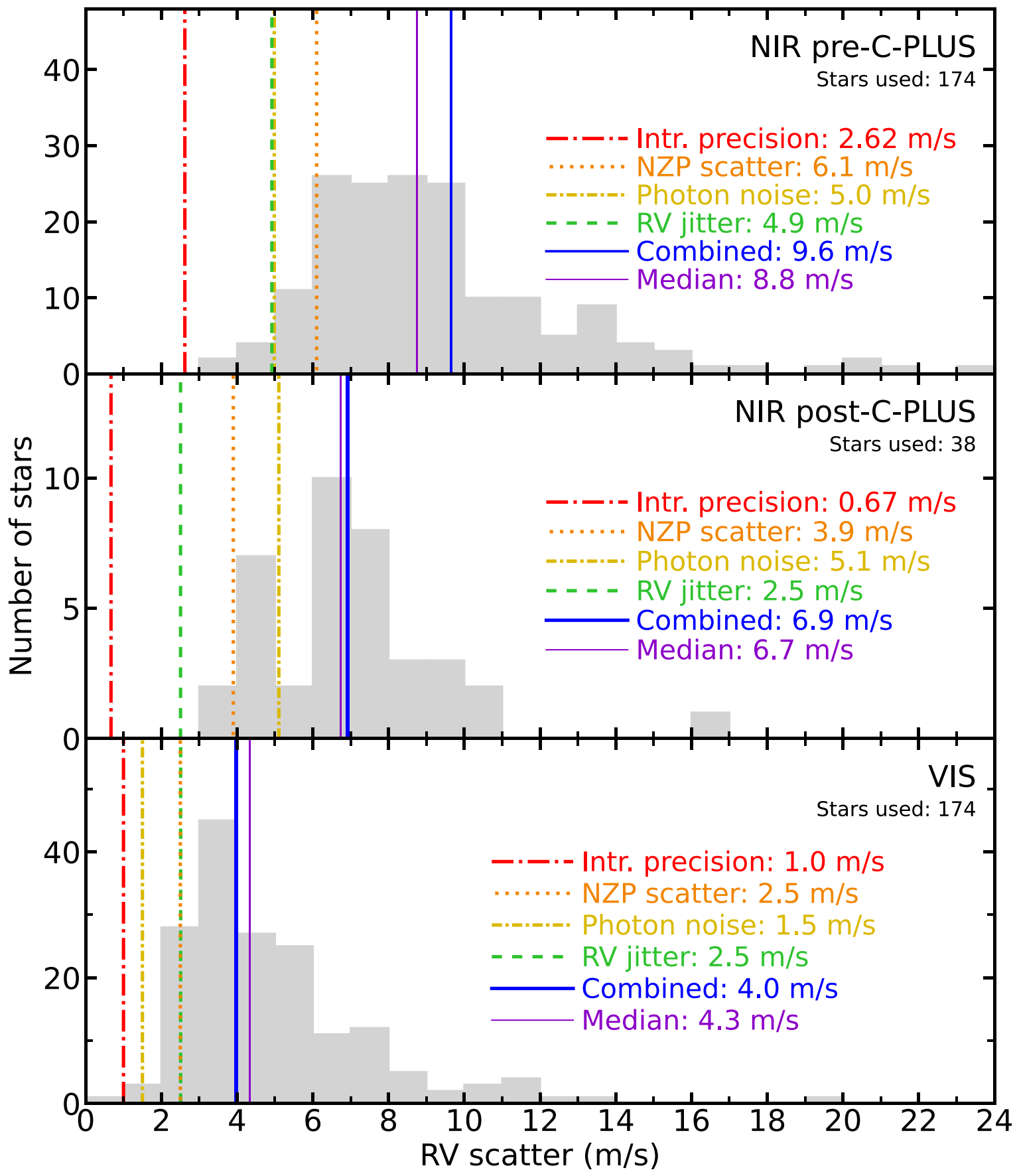

**Fig. 16** Histogram of RV scatter (mlrms) for the three stellar samples: pre-C-PLUS (top), post-C-PLUS (middle), and the CARMENES VIS channel (bottom). Vertical lines represent key RV components: intrinsic RV precision (red dash-dotted), NZP scatter (orange dotted), median photon noise (yellow dash-dotted), median stellar RV jitter (green dashed), the quadratically summed combination (blue thick solid), and the median RV scatter (purple thin solid). After the C-PLUS upgrades, all values–except photon noise–decrease, leading to a noticeable improvement in total RV scatter, although the post-C-PLUS NIR still does not reach the VIS channel performance

A comparison of the pre- and post-C-PLUS distributions reveals a clear improvement in the combined RV scatter, decreasing from 9.6 m s$^{-1}$ to 6.9 m s$^{-1}$. The pre-C-PLUS histogram exhibits a more extended right tail, indicating a greater number of stars with elevated RV variability. While the photon noise remains nearly unchanged between the two samples (5.0 m s$^{-1}$ and 5.1 m s$^{-1}$, respectively), all other contributing terms show a decrease after the C-PLUS upgrades. The evolution of the intrinsic RV precision and the NZP scatter is discussed in Sections 5.1 and 5.2, respectively.

In our analysis, the stellar RV jitter is not a directly measured physical quantity, but a derived one, estimated from the observed RV scatter and the internal error budget, following the method described in [42]. It represents the excess variance beyond the formal uncertainties, but does not directly measure the intrinsic physical variability of the star. Thus, it can also reflect other effects not explicitly accounted for, such as residual instrumental instability, the size and variability of the stellar sample, the number of observations per target, or true long-term spectral variability between the two observing epochs. In the pre-C-PLUS sample, the median stellar jitter is 4.9 m $s^{-1}$, while in the post-C-PLUS sample it drops to 2.5 m $s^{-1}$. This reduction may reflect a combination of improved instrument stability, more conservative internal errors, and the statistical limitations of smaller sample sizes. As noted earlier, jitter estimates can become negative when internal errors are overestimated, directly affecting the median value, which is particularly relevant in the post-C-PLUS dataset.

It is also worth noting that the pre-C-PLUS sample includes a larger number of targets, which increases the likelihood of including stars with higher RV jitter values. This broader sampling may bias the median upward compared to the smaller post-C-PLUS sample.

However, even when we restrict the analysis to stars observed both before and after the upgrade, we still observe a lower median jitter in the post-C-PLUS data. This suggests that the observed reduction is not solely due to sample size, but likely reflects a combination of improved instrumental stability, more conservative internal error estimates, and potential changes in stellar variability between the two epochs.

For the VIS channel, the combined RV scatter is 4.3 m $s^{-1}$, compared to 6.9 m $s^{-1}$ for the post-C-PLUS NIR sample. Both distributions show similarly short right tails, indicating that stars with unusually high RV variability are rare in both cases. The main difference between the VIS and NIR post-C-PLUS datasets arises from photon noise: while the intrinsic RV precision and NZP scatter are comparable, the VIS spectra contain more RV information, resulting in lower photon-limited uncertainties [19]. For the stellar jitter component, a lower value is expected in the NIR due to the reduced impact of chromatic effects at longer wavelengths [29, 43, 44]. In our data, the median stellar jitter is 2.5 m $s^{-1}$ in both VIS and post-C-PLUS NIR samples, and remains unchanged when considering only stars observed in both datasets.

[23] reported a combined RV scatter of 3.6 m $s^{-1}$ for the VIS and 8.4 m $s^{-1}$ for the NIR, based on data that overlaps with the pre-C-PLUS period. In their decomposition, the instrument's jitter contributions were 1.2 m $s^{-1}$ (VIS) and 3.7 m $s^{-1}$ (NIR), the median photon noise values were 1.5 m $s^{-1}$ and 6.9 m $s^{-1}$, and the stellar activity component was assumed to be 3.0 m $s^{-1}$ in both cases. These values are broadly consistent with those obtained in our analysis, and the relative contributions of each noise source show a similar pattern. Differences between the two studies likely stem from the use of different stellar samples, datasets, and RV products. In particular, [23] used RVs that were corrected for NZPs, while our analysis retains the NZP contribution explicitly. Minor discrepancies may also result from differences in how RV scatter is quantified mathematically.

## 6 Conclusion

CARMENES is a dual-channel, high-resolution spectrograph on the 3.5 m Calar Alto telescope, with a visible arm (520–960 nm) and a near-infrared (NIR) arm (960–1710 nm). The NIR channel is cooled to $\sim 140$ K using a nitrogen-gas flow to ensure ther-mal stability. From 2016 to 2020, the VIS showed an instrument radial velocity (RV) precision of 1.2 m $s^{-1}$ and the NIR of 3.7 m $s^{-1}$ [23].

To address the limitations of the NIR channel, the CARMENES-PLUS (C-PLUS) project was launched to design and implement a series of upgrades across the instrument, with a particular focus on improving the thermal stability of the NIR cooling system. The goal was to enhance the intrinsic RV precision by stabilizing the instrument's cooling system.

In this study, we provided an overview of the NIR channel cooling system and described the main upgrades implemented under the C-PLUS project. We analysed calibration RVs recorded during regular operations, along with internal sensor data, to evaluate the effects of the upgrades on the NIR channel's performance, specifically in terms of pressure, temperature, and RV stability. We then compared the intrinsic RV precision of the upgraded NIR channel to its pre-upgrade performance and to that of the VIS channel. We also analysed the scatter of the nightly zero points (NZPs) for the pre-C-PLUS, post-C-PLUS, and VIS datasets, using a maximum-likelihood root-mean-square method (mlrms) to quantify the variability. Finally, we estimated the RV scatter of slowly rotating M dwarfs ($v \sin i_\star \leq 2$ km $s^{-1}$) using the mlrms method. We decomposed the total scatter into four main contributors: the instrument's intrinsic RV precision, the NZP scatter, the median photon noise, and the median stellar RV jitter.

Our results demonstrate that the CARMENES-PLUS upgrades significantly improved the thermal stability of the NIR cooling system, leading to a substantial gain in intrinsic RV precision. We define this precision as the root-mean-square of the residuals after fitting a quadratic model to the Fabry-Pérot calibration data. Following the upgrades, the NIR channel's precision improved from 2.62 m $s^{-1}$ to 0.67 m $s^{-1}$, surpassing the 1.0 m $s^{-1}$ level achieved in the VIS channel. The scatter in the NIR NZPs decreased from 6.1 m $s^{-1}$ before the upgrades to 3.9 m $s^{-1}$ after C-PLUS. For comparison, the VIS channel exhibits a lower NZP scatter of 2.5 m $s^{-1}$.

The median RV scatter of slowly rotating stars in the NIR sample improved from 8.8 m $s^{-1}$ before the upgrades to 6.7 m $s^{-1}$ afterward. In comparison, the VIS sample shows a lower median scatter of 4.3 m $s^{-1}$, primarily due to its lower photon noise (1.5 m $s^{-1}$ in VIS and 5.1 m $s^{-1}$ in NIR). The main drivers of the improvement in the NIR RV scatter are the enhanced intrinsic RV precision and reduced NZP variability, both achieved through the CARMENES-PLUS interventions.

These results demonstrate that the CARMENES-PLUS upgrades have significantly improved the CARMENES NIR channel's performance, both in terms of thermal and radial velocity stability. The upgraded system now enables more reliable and precise RV measurements in the near-infrared, reinforcing the NIR arm as a powerful tool for exoplanet searches around low-mass stars.

**Acknowledgements** CARMENES is an instrument at the Centro Astronómico Hispano en Andalucía (CAHA) at Calar Alto (Almería, Spain), operated jointly by the Junta de Andalucía and the Instituto de Astrofísica de Andalucía (CSIC). CARMENES was funded by the Max-Planck-Gesellschaft (MPG), the Consejo Superior de Investigaciones Científicas (CSIC), the Ministerio de Economía y Competitividad (MINECO) and the European Regional Development Fund (ERDF) through projects FICTS-2011-02, ICTS-2017-07-CAHA-4, and CAHA16-CE-3978, and the members of the CARMENES Consortium (Max-Planck-Institut für Astronomie, Instituto de Astrofísica de Andalucía, Landessternwarte Königstuhl,

Institut de Ciències de l'Espai, Institut für Astrophysik Göttingen, Universidad Complutense de Madrid, Thüringer Landessternwarte Tautenburg, Instituto de Astrofísica de Canarias, Hamburger Sternwarte, Centro de Astrobiología, and Centro Astronómico Hispano-Alemán), with additional contributions by the MINECO, the Deutsche Forschungsgemeinschaft through the Major Research Instrumentation Programme and Research Unit FOR2544 "Blue Planets around Red Stars", the Klaus Tschira Stiftung, the states of Baden-Württemberg and Niedersachsen, and by the Junta de Andalucía. We acknowledge financial support from the Agencia Estatal de Investigación (AEI/10.13039/501100011033) of the Ministerio de Ciencia e Innovación and the ERDF "A way of making Europe" through projects PID2022-137241NB-C4[1:4], PID2021-125627OB-C31, PID2019-109522GB-C52, the Centre of Excellence "Severo Ochoa" and "María de Maeztu" awards to the Instituto de Astrofísica de Canarias (CEX2019-000920-S), Instituto de Astrofísica de Andalucía (CEX2021-001131-S funded by MCIN/AEI/10.13039/501100011033) and Institut de Ciències de l'Espai (CEX2020-001058-M), and the project AST22_00001_8 of the Junta de Andalucía and the Ministerio de Ciencia, Innovación y Universidades funded by the NextGenerationEU and the Plan de Recuperación, Transformación y Resiliencia. We also thank Air Liquide España for the technical assistance during the installation and tests of some C-PLUS upgrades.

**Author Contributions** R.V., R.C.O. and P.J.A. wrote the main manuscript text and developed the figures. R.V., R.C.O. and S.B. contributed to the engineering design of the upgrades and technical tasks. R.V. and P.J.A. performed the radial velocity analysis and gave the scientific background. H.R. contributed to the stellar RV jitter estimation and analysis. M.A., L.H., H.M.M., S.R., D.M.F., J.H., A.L.H., D.B., J.F.L., M.P., J.A.G., J.G.F., J.M., F.H. and J.A. contributed to the engineering design, implemented the upgrades and take care of the maintenance and operation of the instrument. J.A.C, A.K., R.J.M., A.Q., A.R., I.R., W.S. and M.Z. lead the CARMENES consortium. All authors reviewed the manuscript.

**Funding** Open Access funding provided thanks to the CRUE-CSIC agreement with Springer Nature.

**Data Availability** Data is provided within the manuscript or supplementary information files

## Declarations

**Competing interests** The authors declare no competing interests.

## Authors and Affiliations

**R. Varas[1] · R. Calvo-Ortega[1] · P. J. Amado[1] · S. Becerril[2] · H. Ruh[3] · M. Azzaro[4] · L. Hernández[4] · H. Magán-Madinabeitia[1,4] · S. Reinhart[4] · D. Maroto-Fernández[4] · J. Helmling[4] · Á. L. Huelmo[1] · D. Benítez[4] · J. F. López[4] · M. Pineda[4] · J. A. García[4] · J. García de la Fuente[4] · J. Marín[4] · F. Hernández[4] · J. Aceituno[4] · J. A. Caballero[5] · A. Kaminski[6] · R. J. Mathar[7] · A. Quirrenbach[6] · A. Reiners[3] · I. Ribas[8,9] · W. Seifert[6] · M. Zechmeister[3]**

✉ R. Varas
rvaras@iaa.es

1 Instituto de Astrofísica de Andalucía (IAA-CSIC), Glorieta de la Astronomía s/n, Granada 18008, Spain

2 Consorcio IFMIF-DONES ESPAÑA, c/ Gran Vía de Colón 48, Granada 18010, Spain

3 Institut für Astrophysik, Georg-August-Universität, Friedrich-Hund-Platz 1, Göttingen 37077, Germany

4 Centro Astronómico Hispano en Andalucía de Calar Alto (CSIC-Junta de Andalucía), Sierra de los Filabres, Gérgal 04550, Almería, Spain

5 Centro de Astrobiología, CSIC-INTA, Campus ESAC, Camino Bajo del Castillo s/n, Villanueva de la Cañada 28692, Madrid, Spain

6 Landessternwarte, Zentrum für Astronomie der Universität Heidelberg, Königstuhl 12, Heidelberg 69117, Germany

7 Max-Planck-Institut für Astronomie, Königstuhl 17, Heidelberg 69117, Germany

8 Institut de Ciències de l’Espai (ICE, CSIC), Campus UAB, c/ Can Magrans s/n, Bellaterra 08193, Barcelona, Spain

9 Institut d’Estudis Espacials de Catalunya, C/Gran Capità 2-4, Barcelona 08028, Spain